\documentclass[12pt]{elsarticle}
\usepackage{nomencl}
\usepackage{ifthen}
\renewcommand{\nomgroup}[1]{%
\ifthenelse{\equal{#1}{S}}{\item[\textbf{Subscripts}]}{%
\ifthenelse{\equal{#1}{V}}{\item[\textbf{Variables}]}{%
\ifthenelse{\equal{#1}{S}}{\item[\textbf{sets}]}{}}}
}
\makenomenclature
\usepackage{amssymb}

\usepackage[a4paper]{geometry}  
\usepackage{times} 
\usepackage{xurl} 
\usepackage[italian,english]{babel}
\usepackage{latexsym} 
\usepackage{xcolor}

\makeatletter
\newcommand{\@BIBLABEL}{\@emptybiblabel}
\newcommand{\@emptybiblabel}[1]{}
\usepackage{graphicx}
\usepackage{latexsym}
\usepackage{amssymb}
\usepackage{bm}
\usepackage{amsmath}
\usepackage{epsfig}
\usepackage{caption}
\usepackage{subcaption}
\usepackage{graphicx}
\usepackage{caption}
\usepackage{subcaption}
\usepackage{mwe} 
\usepackage{morefloats}
\usepackage{devanagari}
\usepackage{xcolor}
\usepackage{epstopdf}
\begin{document}
\begin{frontmatter}


\title{A New Higher-Order Super Compact Finite Difference Scheme to Study Three-Dimensional Non-Newtonian Flows }

\renewcommand{\thefootnote}{*}
\makeatletter
\def\ps@pprintTitle{%
  \let\@oddhead\@empty
  \let\@evenhead\@empty
  \let\@oddfoot\@empty
  \let\@evenfoot\@oddfoot
}
\makeatother
\author{\textbf{Ashwani Punia$^{1}$, Rajendra K. Ray$^{2}$}\footnote{Corresponding author : Rajendra K. Ray, rajendra@iitmandi.ac.in} \\
  1,2. School of Mathematical and Statistical Sciences, Indian Institute of Technology Mandi,\\
  Mandi, Himachal Pradesh, 175005, India \\
 {\tt mr.punia11@gmail.com}, {\tt rajendra@iitmandi.ac.in}}
\begin{abstract}
This work introduces a new higher-order accurate super compact (HOSC) finite difference scheme for solving complex unsteady three-dimensional (3D) non-Newtonian fluid flow problems. As per the author's knowledge, the proposed scheme is the first ever developed finite difference scheme to solve three-dimensional non-Newtonian flow problem. Not only that, the proposed method is fourth-order accurate in space variables and second-order accurate in time. Also, the proposed scheme utilizes only seven directly adjacent grid points, at the $(n+1)^{th}$ time level, around which the finite difference discretization is made. The governing equations are solved using a time-marching methodology, and pressure is calculated using a pressure-correction strategy based on the modified artificial compressibility method.
Using the power-law viscosity model, we tackle the benchmark problem of a 3D lid-driven cavity, systematically analyzing the varied rheological behavior of shear-thinning ($n=0.5$), shear-thickening $(n=1.5)$, and Newtonian $(n=1.0)$ fluids across different Reynolds numbers $(Re= 1, 50, 100, 200)$. Both Newtonian and non-Newtonian results are carefully investigated in terms of streamlines, velocity variation, pressure distributions, and viscosity contours, and the computed results are validated with the existing benchmark results. The findings demonstrate excellent agreement with the existing results. This extensive analysis, using the new HOSC scheme, not only increases our understanding of non-Newtonian fluid behavior but also provides a robust foundation for future research and practical applications. Thus our work is genuinely novel and pioneering character.
In essence, this study represents a significant leap forward in computational fluid dynamics, offering a transformative perspective on the behavior of 3D non-Newtonian fluids and paving the way for innovative advancements in fluid mechanics and engineering.\\ 
\end{abstract}
\end{frontmatter}
\section{Introduction}
A non-Newtonian fluid represents a distinct class of fluids identified by their viscosity variability when subjected to external forces. Unlike Newtonian fluids, which strictly follow Newton's law of viscosity and retain a constant viscosity under stress, non-Newtonian fluids display a variety of behaviors. Their viscosity varies with the amount of shear stress, resulting in phenomena such as shear-thinning or pseudoplastic fluids, shear-thickening or dilatant fluids, and Bingham plastics. Furthermore, some Non-Newtonian fluids exhibit time-dependent viscosity, as observed in thixotropic or rheopectic liquids. Many common fluids exhibit non-Newtonian characteristics, including silicone oils, polymers, blood, printer ink, and many more. The numerical analysis of non-Newtonian behaviors holds significance in both practical applications and academic research. In last few decades, the majority of studies have predominantly focused on Newtonian fluids. However, it is important to acknowledge that non-Newtonian fluids are commonly utilized in numerous fields. Despite this, only a few studies have investigated the behavior of power-law fluids in two- and three-dimensional enclosures due to the complexity involved with varying viscosity as a function of shear rate.\\
In recent years, different computational approaches have been used to model the flow of power-law fluids in various geometries. Among these, lid-driven cavity flow is particularly exciting because it can incorporate a wide range of complicated hydrodynamic processes. This comprises recirculation patterns, various vortex shapes, singularities, instabilities, and transitional behavior. Several numerical approaches \cite{Benjamin_1979, Kalita_2002, Sahin_2003, Kalita_2014, Li_2022, Benhamou_2022} are presented to handle Newtonian fluid flow. However, the hydrodynamics of non-Newtonian fluids in cavities has received very little attention. \\
Bell and Surana \cite{Bell_1994} investigated two-dimensional, incompressible power-law fluid flow in a lid-driven cavity utilizing the $p$-version least squares finite element formulation. Neofytou \cite{Neofytou_2005} used a 3rd-order upwind finite volume approach to simulate the flow of power-law fluids, Bingham plastics, Casson fluids, and Quemada fluids in a 2D lid-driven cavity. He confirmed the procedure by comparing the numerical data obtained from previously published publications. Zinani and Frey \cite{Zinani_2008} used the finite element method with Galerkin least squares multi-field approximation to investigate the effect of shear-thickening and thinning on the behavior of power-law fluid flows within a single lid-driven cavity for power-law exponents from $0.25$ to $1.5$ and Reynolds numbers ranging from $50$ to $500$.  Kuhlmann et al. \cite{Kuhlmann_1997} conducted experimental and computational investigations of two- and three-dimensional flows within rectangular cavities with anti-parallel motion of side walls for Newtonian fluids. Their studies demonstrated that the fundamental two-dimensional flow is not necessarily unique, since the vortex shape can alter dramatically depending on the side wall velocities and cavity aspect ratios. They also noticed situations where the flow field transitioned to a three-dimensional condition. They also looked at stability and the production of cellular structures at different aspect ratios and Reynolds numbers. Aharonov and Rothman \cite{Aharonov_1993} proposed a new two-dimensional microscopic model to simulate non-Newtonian fluids. Their research demonstrated the model's ability to handle complicated boundaries and multiphase fluids. Using the new model, they examined non-Newtonian flow in porous media and identified a simple scaling equation that linked flux and force. Marn et al. \cite{Marn_2001} investigated the appropriateness of finite difference and finite volume methods for computing incompressible non-Newtonian flow within a 2D lid-driven cavity. Sullivan et al. \cite{Sullivan_2006} used the Lattice Boltzmann Method (LBM) to model power-law fluid flows across porous media in two and three dimensions.
Boyd et al. \cite{Boyd_2006} developed a second-order accurate LBM model designed for non-Newtonian flows. They tested the method's accuracy against power-law fluid flows in solid pipes. Dhiman and Chhabra \cite{Dhiman_2006} numerically studied the two-dimensional flow of power-law fluids over an isolated unconfined square cylinder in the range of conditions $1 \leq Re \leq 45$ and $0.5 \leq n \leq 2$ by using finite volume approach. Their studies demonstrated that shear-thinning behaviour raises drag above its Newtonian equivalent, while shear-thickening behaviour reduces drag below its Newtonian value. Jin et al. \cite{Jin_2017} produced a 3D Navier-Stokes solution using GPUs with the CUDA programming architecture. They employed the fractional step (finite-element) approach to investigate non-Newtonian flow in a lid-driven cubic cavity, with the power-law (Ostwald-deWaele) model serving as the non-linear stress-strain model. Liu et al. \cite{Liu_2018} used a finite difference approach, specifically the Crank-Nicolson method, to numerically estimate a two-dimensional temporal fractional non-Newtonian fluid model. Their findings showed that the Crank-Nicolson difference scheme is successful in modeling the generalized non-Newtonian fluid diffusion model. Hatič et al. \cite{Hatič_2021} addressed the 2D lid-driven cavity problem for a non-Newtonian power-law shear thinning and thickening fluid using a meshless approach.\\
The previously discussed numerical simulations of 3D non-Newtonian power-law fluids were performed using traditional computational fluid dynamics (CFD) techniques, more specific finite element, and finite volume methods. While conventional numerical approaches have offered useful insights, there is a growing demand for more precise and efficient methodologies, especially in 3D non-Newtonian fluid flow situations. Based on the available research, it is evident that in the context of 3D non-Newtonian scenarios, the finite difference technique has received relatively much less attention compared to other approaches. To the best of author's knowledge, there is no higher-order accurate finite difference scheme is available for studying the 3D non-Newtonian power-law model. This is most likely because, in finite difference approach, it difficult to handle the variable viscosity and pressure trem, specially for 3D non-Newtonian problems. Also, the commonly used iterative solvers, such as Gauss-Seidel and Successive Over-Relaxation (SOR) methods, may not be suitable for solving this highly non-linear system of equations, where the coefficient matrix is not diagonally dominant. Here, advanced iterative solvers like the Bi-Conjugate Gradient Stabilized (BiCGSTAB) method or Generalized Minimal Residual (GMRES) method are more useful. Hence, developing an in-house code to tackle these problems is really a very challenging task for any researcher. It's also important to note that the majority of existing finite difference studies have been confined to 2D simulations and predominantly utilized the streamfunction-vorticity formulation of the Navier-Stokes equations, where pressure is not explicitly considered. In contrast, for 3D Navier-Stokes equations, the inclusion of pressure introduces a system with more variables than equations, as there is no evolution equation explicitly governing pressure. Hence, a specialized strategy for correcting the pressure term becomes necessary, leading to significant additional computational effort. This factor also contributes to the slower pace in the development of 3D finite difference methods for non-Newtonian fluids. In our current research, we employ a pressure-correction strategy based on the modified artificial compressibility method, which is highly efficient and straightforward. 
Higher-order compact (HOC) schemes have already established \cite{Spotz_1995,Kalita_2004,Ray_2010} themselves as one of the premier methods for reproducing complex flow dynamics in 2D simulations. Kalita \cite{Kalita_2014} first developed the super compact higher-order scheme for 3D Newtonian fluids, finding it to be second-order accurate in time and fourth-order accurate in space. Recently, Punia and Ray \cite{Punia_2024} extended this scheme to study natural convection and entropy generation inside a cubic cavity for Newtonian fluids, demonstrating its ability to accurately capture fluid phenomena and heat transfer. Until now, no higher-order finite difference scheme has been developed specifically for studying 3D non-Newtonian power-law fluids.

To achieve a more realistic representation of fluid flow, numerical simulations involving three-dimensional (3D) flows are imperative. Unraveling the complexities of 3D non-Newtonian fluids holds paramount importance due to their diverse applications across industries such as food processing, biomedicine, cosmetics, oil and gas industries, wastewater treatment, and  textile industries. Consequently, this area presents a promising and contemporary field for researchers, offering ample opportunities to develop new highly accurate numerical schemes. The straightforward implementation of finite difference schemes has rendered them a preferred choice for researchers and engineers tackling complex non-Newtonian fluid flow problems. Motivated by the need to comprehend the realistic 3D fluid phenomena of non-Newtonian fluids, we introduce a novel higher-order super compact (HOSC) finite difference scheme capable of accurately capturing non-Newtonian fluid dynamics.  

\section{Problem Description and Solution Procedure}
\label{sec:Problem Description and Discretization of Governing Equations}
\subsection{Problem Description}
\begin{figure}[htbp]
 \centering
 \vspace*{5pt}%
 \hspace*{\fill}%
\begin{subfigure}{0.48\textwidth}     
    \centering
    \includegraphics[width=\textwidth]{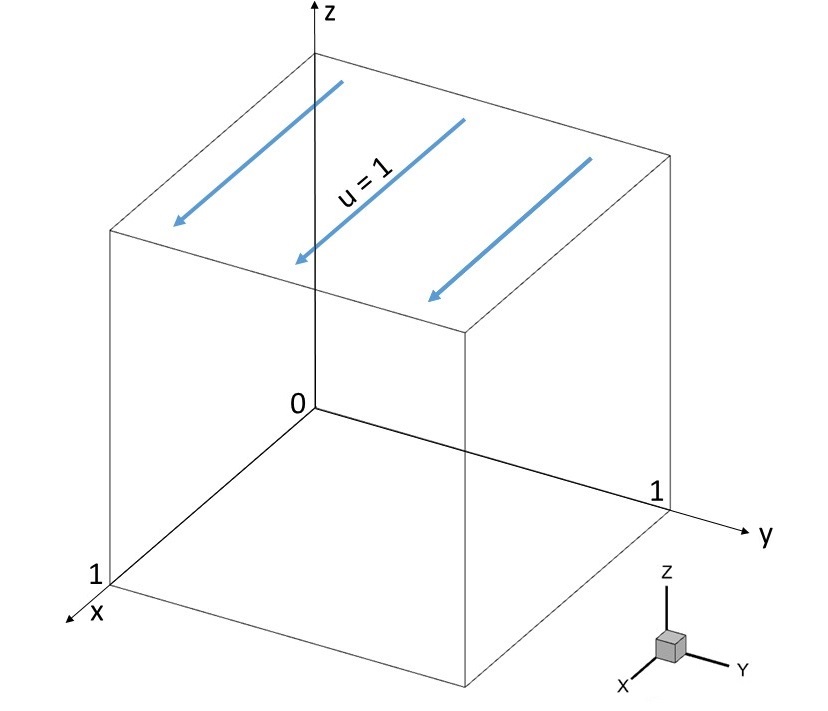}%
    \captionsetup{skip=5pt}%
    \caption{(a)}
    \label{fig:Cavity_3d}
  \end{subfigure}%
 \begin{subfigure}{0.48\textwidth}        
   \centering
    \includegraphics[width=\textwidth]{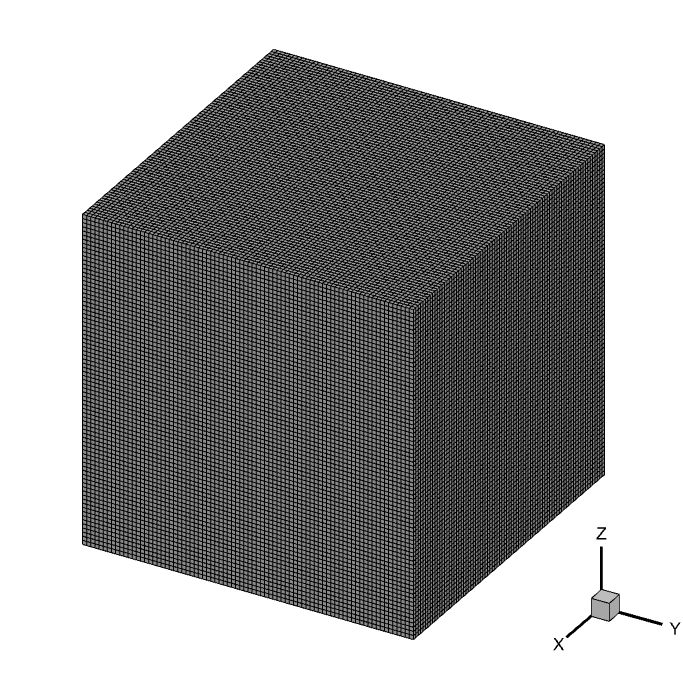}%
    \captionsetup{skip=5pt}%
    \caption{(b)}
    \label{fig:3d_grids}
  \end{subfigure}
  \hspace*{\fill}
  \vspace*{8pt}%
  \hspace*{\fill}%
  \caption{ (a) Illustration of the configuration in the 3D lid-driven cavity scenario and (b) View of the grids at a resolution of $81\times81\times81$.}
  \label{fig:Sche_diag}
\end{figure}

The present study investigates the three-dimensional transient flow of an incompressible non-Newtonian power-law fluid within a closed 3D lid-driven cubic cavity. The flow is induced by the sliding motion of the top wall from left to right at a constant velocity, while the remaining five faces of the cube maintain a stationary (no-slip) condition. A schematic diagram of the problem is shown in Figure \ref{fig:Sche_diag}, along with the grid generation in the Cartesian coordinate system. The dimensionless \cite{Dhiman_2006} form of Cauchy's equations in the $x$-, $y$-, and $z$-components for an incompressible, three-dimensional, transient, and laminar flow are provided below.
\begin{equation}\label{Main_governing_eq_1}
\frac{\partial u}{\partial \tau}+w \frac{\partial u}{\partial z}+v \frac{\partial u}{\partial y}+u \frac{\partial u}{\partial x}=-\frac{\partial p}{\partial x}+\frac{1}{Re} \left[\frac{\partial \tau^{*}_{xx}}{\partial x} + \frac{\partial \tau^{*}_{yx}}{\partial y}  + \frac{\partial \tau^{*}_{zx}}{\partial z}\right] 
\end{equation}

\begin{equation}\label{Main_governing_eq_2}
\frac{\partial v}{\partial \tau}+w \frac{\partial v}{\partial z}+v \frac{\partial v}{\partial y}+u \frac{\partial v}{\partial x}=-\frac{\partial p}{\partial y}+\frac{1}{Re} \left[\frac{\partial \tau^{*}_{xy}}{\partial x} + \frac{\partial \tau^{*}_{yy}}{\partial y}  + \frac{\partial \tau^{*}_{zy}}{\partial z}\right]
\end{equation}

\begin{equation}\label{Main_governing_eq_3}
\frac{\partial w}{\partial \tau}+w \frac{\partial w}{\partial z}+v \frac{\partial w}{\partial y}+u \frac{\partial w}{\partial x}=-\frac{\partial p}{\partial z}+\frac{1}{Re} \left[\frac{\partial \tau^{*}_{xz}}{\partial x} + \frac{\partial \tau^{*}_{yz}}{\partial y}  + \frac{\partial \tau^{*}_{zz}}{\partial z}\right]
\end{equation}
and the continuity equation is expressed as: 
\begin{equation} \label{Main_governing_eq_4} 
\frac{\partial w}{\partial z}+\frac{\partial v}{\partial y}+\frac{\partial u}{\partial x}=0 
\end{equation}
The behavior of the fluid is governed by the power-law model, which is described by the equation:
\begin{equation}\label{Main_governing_eq_5}
\tau^{*}_{ij} = 2 \eta \varepsilon_{i,j}
\end{equation}
where $\varepsilon$ represents the strain rate tensor, which is defined as:
$$
\varepsilon_{i j}=\frac{1}{2}\left(\frac{\partial u_i}{\partial x_j}+\frac{\partial u_j}{\partial x_i}\right)
$$
and $\eta$ represent the effective dynamic viscosity, which is defined as 
\begin{equation} \label{Main_governing_eq_6} 
\eta = I_2^{\frac{n-1}{2}}
\end{equation}
Here, $I_2$ represents the second invariant of the strain rate tensor that measures the intensity of the deformation rate in a fluid. It is crucial for characterizing the flow behavior of non-Newtonian fluids, especially in understanding how their viscosity varies with deformation. It can be expressed as \cite{Zhuang_2018}: 
\begin{equation} \label{Main_governing_eq_7} 
I_2= 2\left[\left(\frac{\partial u}{\partial x}\right)^2+\left(\frac{\partial v}{\partial y}\right)^2+\left(\frac{\partial w}{\partial z}\right)^2 \right]+\left(\frac{\partial u}{\partial y}+\frac{\partial v}{\partial x}\right)^2+\left(\frac{\partial v}{\partial z}+\frac{\partial w}{\partial y}\right)^2 +\left(\frac{\partial w}{\partial x}+\frac{\partial u}{\partial z}\right)^2
\end{equation}
For the power-law viscosity model, the Reynolds number can be defined as \cite{Mendu_2012,Jin_2017}:
$$
Re=\frac{\rho U^{2-n} L^n}{\eta_0}
$$
where $U$ is the lid velocity, $\rho$ is the fluid density, $L$ is the length of the cube, and $\eta_0$ is the consistency index of viscosity.
 We non-dimensionalize all variables of governing equations and consider flow in a cube with $L = 1$ and $U = 1$. 
After using the continuity equation (\ref{Main_governing_eq_4}) and equations (\ref{Main_governing_eq_5}) - (\ref{Main_governing_eq_7}) into equations (\ref{Main_governing_eq_1}) - (\ref{Main_governing_eq_3}) and expressing the equations in their conservative form, we obtain the following momentum equations: \\

\begin{equation}\label{Main_governing_eq_8}
\begin{aligned}
\frac{\partial u}{\partial \tau}+w \frac{\partial u}{\partial z}+v \frac{\partial u}{\partial y}+u \frac{\partial u}{\partial x}=-\frac{\partial p}{\partial x}+\frac{\eta}{Re} \left(\frac{\partial^2 u}{\partial z^2} +\frac{\partial^2 u}{\partial y^2}+ \frac{\partial^2 u}{\partial x^2}\right) \\ + \frac{2}{Re} \left(\varepsilon_{xx} \frac{\partial \eta }{\partial x} +\varepsilon_{yx} \frac{\partial \eta }{\partial y}  + \varepsilon_{zx} \frac{\partial \eta}{\partial z}\right)
\end{aligned}
\end{equation}

\begin{equation}\label{Main_governing_eq_9}
\begin{aligned}
\frac{\partial v}{\partial \tau}+w \frac{\partial v}{\partial z}+v \frac{\partial v}{\partial y}+u \frac{\partial v}{\partial x}=-\frac{\partial p}{\partial y}+\frac{\eta}{Re} \left(\frac{\partial^2 v}{\partial z^2} +\frac{\partial^2 v}{\partial y^2}+ \frac{\partial^2 v}{\partial x^2}\right)  \\ + \frac{2}{Re} \left(\varepsilon_{xy} \frac{\partial \eta }{\partial x} +\varepsilon_{yy} \frac{\partial \eta }{\partial y}  + \varepsilon_{zy} \frac{\partial \eta}{\partial z}\right)
\end{aligned}
\end{equation}

\begin{equation}\label{Main_governing_eq_10}
\begin{aligned}
\frac{\partial w}{\partial \tau}+w \frac{\partial w}{\partial z}+v \frac{\partial w}{\partial y}+u \frac{\partial w}{\partial x}=-\frac{\partial p}{\partial z}+\frac{\eta}{Re} \left(\frac{\partial^2 w}{\partial z^2} +\frac{\partial^2 w}{\partial y^2}+ \frac{\partial^2 w}{\partial x^2}\right)  \\ + \frac{2}{Re} \left(\varepsilon_{xz} \frac{\partial \eta }{\partial x} +\varepsilon_{yz} \frac{\partial \eta }{\partial y}  + \varepsilon_{zz} \frac{\partial \eta}{\partial z}\right)
\end{aligned}
\end{equation}
The dimensionless initial velocity components in the entire domain are set to zero:
\[
u(x, y, z, \tau=0) = 0, \quad v(x, y, z, \tau=0) = 0, \quad w(x, y, z, \tau=0) = 0
\]
and boundary conditions are as follows:\\
1. On the top face of the cubic cavity, a uniform velocity is prescribed along $x$ directions, creating a lid-driven flow. i.e:
\[
u(x, y, 1, \tau) = 1, \quad v(x, y, 1, \tau) = 0, \quad w(x, y, 1, \tau) = 0
\]
2. No-slip condition on five other faces:
\[
U(0, y, z, \tau) = 0, \quad U(1, y, z, \tau) = 0, \quad U(x, 0, z, \tau) = 0, \quad \] \[ U(x, 1, z, \tau) = 0, \quad U(x, y, 0, \tau) = 0
\]
Where, $U=(u,v,w)$.\\
3. Neumann boundary conditions are applied for the pressure $(p)$. 

\subsection{Discretization of Governing Equations using HOSC scheme}
This section discretizes the previously described nonlinear coupled transport equations (\ref{Main_governing_eq_8})-(\ref{Main_governing_eq_10}) using a uniform mesh spacing. We consider the following general partial differential equation with a transport variable designated as ``$\varphi$" inside a continuous domain. 
\begin{equation}
\begin{aligned}
& L(x, y, z,\tau)\frac{\partial \varphi}{\partial\tau}+M^{*}(x, y, z,\tau) \frac{\partial \varphi}{\partial x}+N^{*}(x, y, z,\tau) \frac{\partial \varphi}{\partial y}+O^{*}(x, y, z,\tau) \frac{\partial \varphi}{\partial z}+K^{*}(x, y, z,\tau) \varphi \\
& \quad=\nabla^2 \varphi+F^{*}(x, y, z,\tau),
\end{aligned}\label{eq1}
\end{equation}
The equation includes several variable coefficients ``$L$", ``$M^{*}$", ``$N^{*}$", ``$O^{*}$", ``$K^{*}$" and ``$F^{*}$". This equation represents the convection-diffusion process of numerous fluid variables, such as vorticity, heat, mass, and energy over a continuous domain. In the case of a Newtonian fluid, where viscosity remains constant, treating $L$ as constant suffices. However, for non-Newtonian fluids, where viscosity varies, a constant $L$ is insufficient. In this scenario, $L$ becomes a function of viscosity, which makes the equation extremely challenging. Consequently, to analyze non-Newtonian fluids effectively, we must consider $L$ as a function of $x$, $y$, $z$, and $\tau$. By appropriately selecting the values of ``$L, M^{*}, N^{*}, O^{*}, K^{*}$", and ``$F^{*}$", the equation can effectively represent the momentum equations for Non-Newtonian power-law fluid. As a result, this equation serves as an overall structure capable of representing a wide range of fluid dynamics processes within a single mathematical construct. To ensure a well-defined and physically relevant problem formulation, appropriate boundary conditions for the domain must be established. To discretize the cubical problem domain, we use uniform mesh with increments $h$, $k$, and $l$ for the $x$, $y$, and $z$-direction, respectively. In the discretization process, we first employ the Forward-Time Central-Space (FTCS) scheme to equation (\ref{eq1}). Using the FTCS approximation, we can approximate Eq. (\ref{eq1}) at the general node $(i, j, k)$ as follows:
\begin{equation} \begin{aligned}
\left(L \delta_\tau^{+}+O^{*}\delta_z+N^{*}\delta_y+M^{*} \delta_x-\delta_z^2-\delta_y^2-\delta_x^2 + K^{*}\right) \varphi_{i j k}-\xi_{i j k}=F^{*}_{i j k},
\end{aligned} \label{eq2_} \end{equation} 
In the above equation, $\varphi_{ijk}$ represents the functional value of transport variable $\varphi$ at a three-dimensional grid point $(x_i, y_j, z_k)$. The operator $\delta_x, \delta_x^2, \delta_y, \delta_y^2, \delta_z$ and $\delta_z^2$ are associated with first and second-order central differences along the space variables $x, y, z,$ respectively, and $\delta_\tau^{+}$ is the first order forward difference along the time variable. Truncation error $\xi_{i j k}$ associated with the present numerical method, employing a uniform time step $\Delta\tau$, serves as a measure of the introduced error due to the discretization process, as described by:
\begin{equation}
\begin{aligned}
\xi_{i j k}= & {\left[L \frac{\Delta\tau}{2} \frac{\partial^2 \varphi}{\partial\tau^2}-\frac{k^2}{12}\left(\frac{\partial^4 \varphi}{\partial y^4} - 2 N^{*} \frac{\partial^3 \varphi}{\partial y^3}\right)-\frac{h^2}{12}\left(\frac{\partial^4 \varphi}{\partial x^4}-2 M^{*} \frac{\partial^3 \varphi}{\partial x^3}\right)\right.} \\
& \left.-\frac{l^2}{12}\left(\frac{\partial^4 \varphi}{\partial z^4}-2 O^{*} \frac{\partial^3 \varphi}{\partial z^3}\right)\right]_{i j k}+O\left(\Delta\tau^2, h^4, k^4, l^4\right).
\end{aligned}\label{eq3}
\end{equation}
To achieve higher temporal and spatial accuracy of equation (\ref{eq1}), a compact approximation for the derivatives of the leading term in equation (\ref{eq3}) is used. This approach leads to a formulation characterized by reduced truncation error.
To fulfill this objective, Eq. (\ref{eq1}) is treated as an auxiliary relationship for calculating higher-order derivatives, which means higher derivatives ($3^{rd}$ and $4^{th}$ with respect to space and second derivative with respect to time) are computed from Eq. (\ref{eq1}). 
For instance, to calculate the second derivative with respect to time, the backward temporal difference method is applied to the variables $L, M^{*}, N^{*}, O^{*}, K^{*},$ and $F^{*}$ and forward difference method is employed for the transport variable $\varphi$ \cite{Kalita_2014}. 
This enables the representation of derivatives in the initial right-hand sided term of Eq. (\ref{eq3}) in the following manner:
\begin{equation}
\begin{aligned}
\left.L_{i j k} \frac{\partial^2 \varphi}{\partial t^2}\right|_{i j k}= & \left(\delta_z^2+\delta_y^2+\delta_x^2-K^{*}_{i j k}-M^{*}_{i j k} \delta_x-N^{*}_{i j k} \delta_y-O^{*}_{i j k} \delta_z\right) \delta_\tau^{+} \varphi_{i j k} \\
& -\left(\delta_\tau^{-} M^{*}_{i j k} \delta_x+\delta_\tau^{-} N^{*}_{i j k} \delta_y+\delta_\tau^{-} K^{*}_{i j k}+\delta_\tau^{-} O^{*}_{i j k} \delta_z \right)\varphi_{i j k} +\delta_\tau^{-} F^{*}_{i j k}\\
& \delta_\tau^{-} L_{i j k} \delta_\tau^{+} \varphi_{i j k} +O\left(\Delta\tau, h^2, k^2, l^2\right),
\end{aligned} \label{eq4}
\end{equation}
The operators $\delta_\tau^{+}$ and $\delta_\tau^{-}$ represent the first-order forward and backward difference with regard to the time, respectively. 
The operators $\delta_x, \delta_x^2, \delta_y, \delta_y^2, \delta_z$ and $\delta_z^2$ are associated with first and second-order central differences along the spatial variables. 
Similarly, the other derivatives in (\ref{eq3}) can be determined using (\ref{eq1}). By substituting these derivatives into Eq. (\ref{eq3}) and then replacing 
$\xi_{i j k}$ in Eq. (\ref{eq2_}), we derive the following estimation, achieving an order of accuracy $O\left(\Delta \tau^2, h^4, k^4, l^4\right)$ for the primary governing equation (\ref{eq1}).

$$
\begin{aligned}
L_{i j k} \bigg[1 & +\left(\frac{k^2}{12}-\frac{\Delta\tau}{2 L_{i j k}}\right)\left(\delta_y^2-N^{*}_{i j k} \delta_y\right) +\left(\frac{h^2}{12}-\frac{\Delta\tau}{2 L_{i j k}}\right)\left(\delta_x^2-M^{*}_{i j k} \delta_x\right) \\
& +\left(\frac{l^2}{12}-\frac{\Delta\tau}{2 L_{i j k}}\right)\left(\delta_z^2-O^{*}_{i j k} \delta_z\right) \\
&  + \frac{\Delta\tau}{2 L_{i j k}} \left( K^{*}_{i j k} + \delta_\tau^{-} L_{i j k} +  \frac{h^2}{6\Delta\tau}\delta_x^2 L_{i j k} + \frac{k^2}{6\Delta\tau}\delta_y^2 L_{i j k}
+ \frac{l^2}{6\Delta\tau}\delta_z^2 L_{i j k}\right)\\
&  +\frac{\Delta\tau}{2 L_{i j k}} \left(- \frac{h^2}{6\Delta\tau}M^{*}_{i j k}\delta_x L_{i j k}
- \frac{k^2}{6\Delta\tau}N^{*}_{i j k}\delta_y L_{i j k}
- \frac{l^2}{6\Delta\tau}O^{*}_{i j k}\delta_z L_{i j k}\right)\\
& + \frac{\Delta\tau}{2 L_{i j k}} \left(
 (\frac{h^2}{6 L_{i j k}}\delta_x L_{i j k})\delta_x
+ (\frac{k^2}{6 L_{i j k}}\delta_y L_{i j k})\delta_y
+ (\frac{l^2}{6 L_{i j k}}\delta_z L_{i j k})\delta_z\right) \bigg]\delta_\tau^{+} \varphi_{i j k} \\
& +\left(-\alpha_{i j k} \delta_x^2-\beta_{i j k} \delta_y^2-\gamma_{i j k} \delta_z^2+M1_{i j k} \delta_x+M2_{i j k} \delta_y+M3_{i j k} \delta_z+M4_{i j k}\right) \varphi_{i j k} \\
& -\frac{h^2+k^2}{12}\left(\delta_x^2 \delta_y^2-M^{*}_{i j k} \delta_x \delta_y^2-N^{*}_{i j k} \delta_x^2 \delta_y-p1_{i j k} \delta_x \delta_y\right) \varphi_{i j k}
\end{aligned}
$$
\begin{equation}
\begin{aligned}
& -\frac{k^2+l^2}{12}\left(\delta_y^2 \delta_z^2-N^{*}_{i j k} \delta_y \delta_z^2-O^{*}_{i j k} \delta_y^2 \delta_z-q1_{i j k} \delta_y \delta_z\right) \varphi_{i j k} \\
& -\frac{l^2+h^2}{12}\left(\delta_z^2 \delta_x^2-O^{*}_{i j k} \delta_z \delta_x^2-M^{*}_{i j k} \delta_z^2 \delta_x-r1_{i j k} \delta_z \delta_x\right) \varphi_{i j k} = R_{i j k}\\
\end{aligned}\label{eq5}
\end{equation}
The coefficients $\alpha_{i j k}, \beta_{i j k}, \gamma_{i j k}, M1_{i j k}, M2_{i j k}, M3_{i j k}, M4_{i j k}, R_{i j k}, p1_{i j k}, q1_{i j k}$ and $r1_{i j k}$ are as follows:
$$
\begin{aligned}
& \alpha_{i j k}=\frac{h^2}{12}\left({M^{*}}_{i j k}^2-{K^{*}}_{i j k}-2 \delta_x {M^{*}}_{i j k}\right)+1 \text {, } \\
& \beta_{i j k}=\frac{k^2}{12}\left({N^{*}}_{i j k}^2-{K^{*}}_{i j k}-2 \delta_y {N^{*}}_{i j k}\right)+1 \text {, } \\
& \gamma_{i j k}=\frac{l^2}{12}\left({O^{*}}_{i j k}^2-{K^{*}}_{i j k}-2 \delta_z {O^{*}}_{i j k}\right)+1\text {, } \\
& M1_{i j k}=\left[\frac{h^2}{12}\left(\delta_x^2-{M^{*}}_{i j k} \delta_x\right)+\frac{k^2}{12}\left(\delta_y^2-{N^{*}}_{i j k} \delta_y\right)+\frac{l^2}{12}\left(\delta_z^2-{O^{*}}_{i j k} \delta_z\right)+\frac{\Delta\tau}{2} \delta_\tau^{-}+1\right] {M^{*}}_{i j k} \\
& -\frac{h^2}{12}\left({M^{*}}_{i j k}-2 \delta_x\right) {K^{*}}_{i j k}, \\
& M2_{i j k}=\left[\frac{h^2}{12}\left(\delta_x^2-{M^{*}}_{i j k} \delta_x\right)+\frac{k^2}{12}\left(\delta_y^2-{N^{*}}_{i j k} \delta_y\right)+\frac{l^2}{12}\left(\delta_z^2-{O^{*}}_{i j k} \delta_z\right)+\frac{\Delta\tau}{2} \delta_\tau^{-}+1\right] {N^{*}}_{i j k} \\
& -\frac{k^2}{12}\left({N^{*}}_{i j k}-2 \delta_y\right) {K^{*}}_{i j k} \text {, } \\
& M3_{i j k}=\left[\frac{h^2}{12}\left(\delta_x^2-{M^{*}}_{i j k} \delta_x\right)+\frac{k^2}{12}\left(\delta_y^2-{N^{*}}_{i j k} \delta_y\right)+\frac{l^2}{12}\left(\delta_z^2-{O^{*}}_{i j k} \delta_z\right)+\frac{\Delta\tau}{2} \delta_\tau^{-}+1\right] {O^{*}}_{i j k} \\
& -\frac{l^2}{12}\left({O^{*}}_{i j k}-2 \delta_z\right) {K^{*}}_{i j k} \text {, } \\
& M4_{i j k}=\left[\frac{h^2}{12}\left(\delta_x^2-{M^{*}}_{i j k} \delta_x\right)+\frac{k^2}{12}\left(\delta_y^2-{N^{*}}_{i j k} \delta_y\right)+\frac{l^2}{12}\left(\delta_z^2-{O^{*}}_{i j k} \delta_z\right)+\frac{\Delta\tau}{2} \delta_\tau^{-}+1\right] {K^{*}}_{i j k}, \\
& R_{i j k}=\left[\frac{h^2}{12}\left(\delta_x^2-{M^{*}}_{i j k} \delta_x\right)+\frac{k^2}{12}\left(\delta_y^2-{N^{*}}_{i j k} \delta_y\right)+\frac{l^2}{12}\left(\delta_z^2-{O^{*}}_{i j k} \delta_z\right)+\frac{\Delta\tau}{2} \delta_\tau^{-}+1\right] {F^{*}}_{i j k}, \\
& p1_{i j k}=-{M^{*}}_{i j k} {N^{*}}_{i j k}+\frac{2}{h^2+k^2}\left(k^2 \delta_y {M^{*}}_{i j k}+h^2 \delta_x {N^{*}}_{i j k}\right) \text {, } \\
& q1_{i j k}=-{N^{*}}_{i j k} {O^{*}}_{i j k}+\frac{2}{k^2+l^2}\left(l^2 \delta_z {N^{*}}_{i j k}+k^2 \delta_y {O^{*}}_{i j k}\right) \text {, } \\
& r1_{i j k}=-{O^{*}}_{i j k} {M^{*}}_{i j k}+\frac{2}{l^2+h^2}\left(h^2 \delta_x {O^{*}}_{i j k}+l^2 \delta_z {M^{*}}_{i j k}\right) . \\
&
\end{aligned}
$$

In simpler terms, Equation (\ref{eq5}) can be expressed as follows:
\begin{equation}
\begin{aligned}
A_1\varphi_{i+1 j k}^{n+1}+A_2\varphi_{i-1 j k}^{n+1}+A_3\varphi_{i j+1 k}^{n+1}+A_4\varphi_{i j-1 k}^{n+1}+A_5\varphi_{i j k+1}^{n+1}+A_6\varphi_{i j k-1}^{n+1}+A_7\varphi_{i j k}^{n+1}=B_1\varphi_{i+1 j k}^{n}
\\+B_2\varphi_{i j k}^{n}+B_3\varphi_{i-1 j k}^{n}+B_4\varphi_{i j+1 k}^{n}+B_5\varphi_{i j-1 k}^{n}+B_6\varphi_{i j k+1}^{n}+ B_7\varphi_{i j k-1}^{n}+B_8\varphi_{i+1 j+1 k}^{n} \quad \quad  \\
+B_9\varphi_{i-1 j+1 k}^{n}+B_{10}\varphi_{i+1 j-1 k}^{n}+B_{11}\varphi_{i-1 j-1 k}^{n}+B_{12}\varphi_{i j+1 k+1}^{n}+B_{13}\varphi_{i j-1 k+1}^{n}+B_{14}\varphi_{i j+1 k-1}^{n} \\
+B_{15}\varphi_{i j-1 k-1}^{n}+B_{16}\varphi_{i+1 j k+1}^{n} +B_{17}\varphi_{i+1 j k-1}^{n}+B_{18}\varphi_{i-1 j k+1}^{n}+B_{19}\varphi_{i-1 j k-1}^{n}+\Delta t R_{i,j,k} \quad \quad \quad \quad \quad 
 \end{aligned}\label{eq_implicit}
\end{equation}
where,
$
\begin{aligned}
A_1=L_{i j k}\left(\frac{M_1}{h^2}-\frac{M_1A^{*}_{i j k}}{2h}+\frac{M_{15}}{2h}\right){, } \quad M_1=\left(\frac{h^2}{12}-\frac{\Delta \tau}{2L_{i j k}}\right)\\
\end{aligned}
$

$
\begin{aligned}
A_2=L_{i j k}\left(\frac{M_1}{h^2}+\frac{M_1A^{*}_{i j k}}{2h}+\frac{M_{15}}{2h}\right)\\
\end{aligned}
$

$
\begin{aligned}
A_3=L_{i j k}\left(\frac{M_2}{k^2}-\frac{M_2B^{*}_{i j k}}{2K}+\frac{M_{16}}{2k}\right){, } \quad M_2=\left(\frac{k^2}{12}-\frac{\Delta \tau}{2L_{i j k}}\right)\\
\end{aligned}
$

$
\begin{aligned}
A_4=L_{i j k}\left(\frac{M_2}{k^2}+\frac{M_2B^{*}_{i j k}}{2K}+\frac{M_{16}}{2k}\right)\\
\end{aligned}
$

$
\begin{aligned}
A_5=L_{i j k}\left(\frac{M_3}{l^2}-\frac{M_3C^{*}_{i j k}}{2l}+\frac{M_{17}}{2l}\right){, } \quad M_3=\left(\frac{l^2}{12}-\frac{\Delta \tau}{2L_{i j k}}\right)\\
\end{aligned}
$

$
\begin{aligned}
A_6=L_{i j k}\left(\frac{M_3}{l^2}+\frac{M_3C^{*}_{i j k}}{2l}+\frac{M_{17}}{2l}\right)\\
\end{aligned}
$

$
\begin{aligned}
A_7=L_{i j k}\left(1-\frac{2M_1}{h^2}-\frac{2M_2}{k^2}-\frac{2M_3}{l^2}+ \frac{2M_5}{\Delta \tau} + M_4K^{*}_{i j k}\right) {, } \quad M_4=\left(\frac{\Delta \tau}{2L_{i j k}}\right)\\
\end{aligned}
$

$
\begin{aligned}
B_1=\left(\frac{N1_{i j k}}{h^2}+\frac{N4_{i j k}}{2h}-\frac{2\Delta \tau M_5}{h^2k^2} + \frac{2\Delta \tau M_5A^{*}_{i j k}}{2hk^2} -\frac{2\Delta \tau M_7}{l^2k^2}+ \frac{2\Delta \tau M_7A^{*}_{i j k}}{2hl^2} + L_{i j k}\frac{M_{25}}{2h} \right)\\
\end{aligned}
$

$
\begin{aligned}
B_2=-2\left(\frac{N1_{i j k}}{h^2}+\frac{N2_{i j k}}{k^2}+\frac{N3_{i j k}}{l^2}\right) + \left(N7_{i j k}K^{*}_{i j k}+N8_{i j k}D_{i j k}\right)+ \frac{4\Delta \tau M_5}{h^2k^2} \\+\frac{4\Delta \tau M_6}{k^2l^2}+\frac{4\Delta \tau M_7}{h^2l^2} \quad \quad \quad \quad \quad \quad \quad \quad \quad \quad \quad \quad \quad \quad \quad \quad \quad \quad \quad \quad \\
\end{aligned}
$

$
\begin{aligned}
B_3=\left(\frac{N1_{i j k}}{h^2}+\frac{N4_{i j k}}{2h}-\frac{2\Delta \tau M_5}{h^2k^2} - \frac{2\Delta \tau M_5 A^{*}_{i j k}}{2hk^2} -\frac{2\Delta \tau M_7}{l^2k^2}- \frac{2\Delta \tau M_7A^{*}_{i j k}}{2hl^2}- L_{i j k}\frac{M_{25}}{2h}\right)\\
B_4=\left(\frac{N2_{i j k}}{k^2}+\frac{N5_{i j k}}{2k}-\frac{2\Delta \tau M_5}{h^2k^2} + \frac{2\Delta \tau M_5B^{*}_{i j k}}{2h^2k} -\frac{2\Delta \tau M_6}{l^2k^2}+ \frac{2\Delta \tau M_6B^{*}_{i j k}}{2kl^2} + L_{i j k}\frac{M_{26}}{2k}\right)\\
B_5=\left(\frac{N2_{i j k}}{k^2}-\frac{N5_{i j k}}{2k}-\frac{2\Delta \tau M_5}{h^2k^2} - \frac{2\Delta \tau M_5B^{*}_{i j k}}{2h^2k} -\frac{2\Delta \tau M_6}{l^2k^2}- \frac{2\Delta \tau M_6B^{*}_{i j k}}{2kl^2}- L_{i j k}\frac{M_{26}}{2k}\right)\\
B_6=\left(\frac{N3_{i j k}}{l^2}+\frac{N6_{i j k}}{2l}-\frac{2\Delta \tau M_6}{l^2k^2} + \frac{2\Delta \tau M_6C^{*}_{i j k}}{2k^2l} -\frac{2\Delta \tau M_7}{l^2h^2}+ \frac{2\Delta \tau M_7C^{*}_{i j k}}{2lh^2} + L_{i j k}\frac{M_{27}}{2l}\right)\\
B_7=\left(\frac{N3_{i j k}}{l^2}-\frac{N6_{i j k}}{2l}-\frac{2\Delta \tau M_6}{l^2k^2} - \frac{2\Delta \tau M_6C^{*}_{i j k}}{2k^2l} -\frac{2\Delta \tau M_7}{l^2h^2}- \frac{2\Delta \tau M_7C^{*}_{i j k}}{2lh^2} - L_{i j k}\frac{M_{27}}{2l}\right)\\
\end{aligned}
$

$
\begin{aligned}
&&\\
B_8=\left(\frac{\Delta \tau M_5}{h^2k^2}- \frac{\Delta \tau M_5 A^{*}_{i j k}}{2hk^2}- \frac{\Delta \tau M_5 B^{*}_{i j k}}{2h^2k}-\frac{\Delta \tau M_5 p1^{*}_{i j k}}{4hk}\right)\\
B_9=\left(\frac{\Delta \tau M_5}{h^2k^2}+ \frac{\Delta \tau M_5 A^{*}_{i j k}}{2hk^2}- \frac{\Delta \tau M_5 B^{*}_{i j k}}{2h^2k}+\frac{\Delta \tau M_5 p1^{*}_{i j k}}{4hk}\right)\\
B_{10}=\left(\frac{\Delta \tau M_5}{h^2k^2}- \frac{\Delta \tau M_5 A^{*}_{i j k}}{2hk^2}+ \frac{\Delta \tau M_5 B^{*}_{i j k}}{2h^2k}+\frac{\Delta \tau M_5 p1^{*}_{i j k}}{4hk}\right)\\
B_{11}=\left(\frac{\Delta \tau M_5}{h^2k^2}+ \frac{\Delta \tau M_5 A^{*}_{i j k}}{2hk^2}+ \frac{\Delta \tau M_5 B^{*}_{i j k}}{2h^2k}-\frac{\Delta \tau M_5 p1^{*}_{i j k}}{4hk}\right)\\
B_{12}=\left(\frac{\Delta \tau M_6}{l^2k^2}- \frac{\Delta \tau M_6 B^{*}_{i j k}}{2kl^2}- \frac{\Delta \tau M_6 C^{*}_{i j k}}{2k^2l}-\frac{\Delta \tau M_6 q1^{*}_{i j k}}{4kl}\right)\\
B_{13}=\left(\frac{\Delta \tau M_6}{l^2k^2}+ \frac{\Delta \tau M_6 B^{*}_{i j k}}{2kl^2}- \frac{\Delta \tau M_6 C^{*}_{i j k}}{2k^2l}+\frac{\Delta \tau M_6 q1^{*}_{i j k}}{4kl}\right)\\
B_{14}=\left(\frac{\Delta \tau M_6}{l^2k^2}- \frac{\Delta \tau M_6 B^{*}_{i j k}}{2kl^2}+ \frac{\Delta \tau M_6 C^{*}_{i j k}}{2k^2l}+\frac{\Delta \tau M_6 q1^{*}_{i j k}}{4kl}\right)\\
B_{15}=\left(\frac{\Delta \tau M_6}{l^2k^2}+ \frac{\Delta \tau M_6 B^{*}_{i j k}}{2kl^2}+ \frac{\Delta \tau M_6 C^{*}_{i j k}}{2k^2l}-\frac{\Delta \tau M_6 q1^{*}_{i j k}}{4kl}\right)\\
B_{16}=\left(\frac{\Delta \tau M_7}{l^2h^2}- \frac{\Delta \tau M_7 C^{*}_{i j k}}{2lh^2}- \frac{\Delta \tau M_7 A^{*}_{i j k}}{2l^h}-\frac{\Delta \tau M_7 r1^{*}_{i j k}}{4lh}\right)\\
B_{17}=\left(\frac{\Delta \tau M_7}{l^2h^2}+ \frac{\Delta \tau M_7 C^{*}_{i j k}}{2lh^2}- \frac{\Delta \tau M_7 A^{*}_{i j k}}{2l^h}+\frac{\Delta \tau M_7 r1^{*}_{i j k}}{4lh}\right)\\
B_{18}=\left(\frac{\Delta \tau M_7}{l^2h^2}- \frac{\Delta \tau M_7 C^{*}_{i j k}}{2lh^2}+ \frac{\Delta \tau M_7 A^{*}_{i j k}}{2l^h}+\frac{\Delta \tau M_7 r1^{*}_{i j k}}{4lh}\right)\\
B_{19}=\left(\frac{\Delta \tau M_7}{l^2h^2}+ \frac{\Delta \tau M_7 C^{*}_{i j k}}{2lh^2}+ \frac{\Delta \tau M_7 A^{*}_{i j k}}{2l^h}-\frac{\Delta \tau M_7 r1^{*}_{i j k}}{4lh}\right)\\
\end{aligned}
$

$
\begin{aligned}
M_5=\left(\frac{h^2+k^2}{12}\right), \quad M_6=\left(\frac{k^2+l^2}{12}\right), \quad M_7=\left(\frac{l^2+h^2}{12}\right)
\end{aligned}
$

$
\begin{aligned}
N1_{i,j,k}=\left(LM_1+\Delta \tau \alpha_{i,j,k} \right), \quad N2_{i,j,k}=\left(LM_2+\Delta \tau \beta_{i,j,k} \right)
\end{aligned}
$

$
\begin{aligned}
N3_{i,j,k}=\left(LM_3+\Delta \tau \gamma_{i,j,k} \right), \quad N4_{i,j,k}=\left(-LM_1A^{*}_{i j k}-\Delta \tau A_{i j k} \right)
\end{aligned}
$

$
\begin{aligned}
N5_{i,j,k}=\left(-LM_2B^{*}_{i j k}-\Delta \tau B_{i j k} \right), \quad N6_{i,j,k}=\left(-LM_3C^{*}_{i j k}-\Delta \tau C_{i j k} \right)
\end{aligned}
$

$
\begin{aligned}
N7_{i,j,k}=\left(LM_4\right), \quad N8_{i,j,k}=\left(-\Delta \tau \right) 
\end{aligned}
$

$
\begin{aligned}
M15_{i,j,k}=M25_{i,j,k}=\left(\frac{h^2}{6L_{i j k}}\right)\delta_x {L_{i j k}}
\end{aligned}
$

$
\begin{aligned}
M16_{i,j,k}=M26_{i,j,k}=\left(\frac{k^2}{6L_{i j k}}\right)\delta_y {L_{i j k}}
\end{aligned}
$

$
\begin{aligned}
M17_{i,j,k}=M27_{i,j,k}=\left(\frac{l^2}{6L_{i j k}}\right)\delta_z {L_{i j k}}
\end{aligned}
$

$
\begin{aligned}
& M5_{i j k}=\left[\frac{h^2}{12}\left(\delta_x^2-{M^{*}}_{i j k} \delta_x\right)+\frac{k^2}{12}\left(\delta_y^2-{N^{*}}_{i j k} \delta_y\right)+\frac{l^2}{12}\left(\delta_z^2-{O^{*}}_{i j k} \delta_z\right)+ \frac{\Delta \tau \delta_\tau^{-}}{2} \right] {L}_{i j k},\\
&
\end{aligned}
$

\begin{figure}
    \centering
    \includegraphics[width=0.8\textwidth]{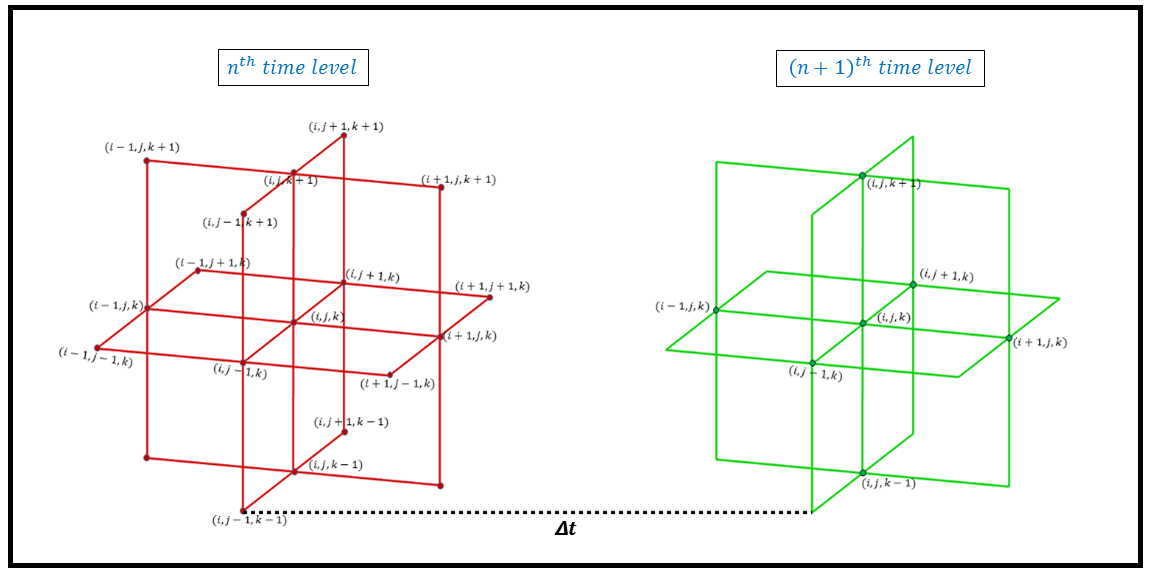}
    \caption{The super-compact unsteady stencil}
    \label{fig:stencil}
\end{figure}

$\\
\begin{aligned}
 \quad
\end{aligned}
$\\
By employing Eq. (\ref{eq5}), an implicit higher-order accurate finite difference method is developed. This is accomplished by utilising a $(19,7)$ stencil, as seen in Figure \ref{fig:stencil}. This approach results in a concise seven-point stencil at the $(n + 1)^{th}$ time step, significantly reducing computational complexity. It’s noteworthy that many high-order compact methods, such as those discussed in \cite{Spotz_1995, Kalita_2002, Ray_2017}, designed even for two-dimensional convection-diffusion equations, typically necessitate a nine-point stencil at the $(n+1)^{th}$ time level. These methods are all developed for the Newtonian model only. However, in the HOSC scheme for the non-Newtonian case, where variable viscosity makes the equations very complex in comparison to the Newtonian case, the necessity is diminished to just a seven-point stencil at the $(n + 1)^{th}$ time level, even for three-dimensional scenarios. This scheme offers dual benefits. Firstly, it simplifies to a seven-point stencil, requiring only the $(i, j, k)$ $^{th}$ point and its six adjacent points (as illustrated in Figure \ref{fig:stencil}) at the $(n + 1)^{th}$ time level. Secondly, it eliminates the necessity for extensive corner points and significantly reduces the number of points required for the approximation, thereby enhancing computational efficiency.
In this discretization process of the momentum equations (Eqs. (\ref{Main_governing_eq_8}) - (\ref{Main_governing_eq_10})), we assign the transport variable ``$\varphi$" as $u$, $v$ and $w$ for the x-,y- and z-momentum equations, respectively. The coefficients are then specified as follows:\\
$L$ is set to ($Re/ \eta$).\\
$M^{*}$ is set to $(Re \cdot u)/ \eta $, where $u$ represents the velocity component in the $x$ direction.\\
$N^{*}$ is set to $(Re \cdot v)/ \eta $, where $v$ represents the velocity component in the $y$ direction.\\
$O^{*}$ is set to $(Re \cdot w)/ \eta $, where $w$ represents the velocity component in the $z$ direction.\\
$K^{*}$ is assigned a value of 0.\\
$F^{*}$ represents one of three options:  -$\frac{ Re}{ \eta}\frac{\partial p}{\partial x}$ +  $\frac{2}{\eta} \left(\varepsilon_{xx} \frac{\partial \eta }{\partial x} +\varepsilon_{yx} \frac{\partial \eta }{\partial y}  + \varepsilon_{zx} \frac{\partial \eta}{\partial z}\right)$, or
-$\frac{Re}{ \eta}\frac{\partial p}{\partial y}$ +  $\frac{2}{\eta} \left(\varepsilon_{xy} \frac{\partial \eta }{\partial x} +\varepsilon_{yy} \frac{\partial \eta }{\partial y}  + \varepsilon_{zy} \frac{\partial \eta}{\partial z}\right)$, or 
-$\frac{Re}{\eta}\frac{\partial p}{\partial z}$ +  $\frac{2}{\eta} \left(\varepsilon_{xz} \frac{\partial \eta }{\partial x} +\varepsilon_{yz} \frac{\partial \eta }{\partial y}  + \varepsilon_{zz} \frac{\partial \eta}{\partial z}\right)$, depending on the particular momentum equation.
In the absence of recognised analytical formulations for pressure, we rely on numerical approximations for pressure gradients. This is carried out by using central difference estimates at the domain's inner grid points and typical one-sided approximations at the boundary nodes. Applying the HOSC method to the governing equations (Eqs. (\ref{Main_governing_eq_8}) - (\ref{Main_governing_eq_10})) at all grid points results in a system of algebraic equations with an asymmetric sparse coefficient matrix that lacks diagonal dominance. As a result, common iterative approaches such as Gauss-Seidel and SOR are unsuccessful in this setting. To solve the system of algebraic equations, a Hybrid Biconjugate Gradient Stabilised approach is used without any preconditioning \cite{Kelley_1995, Spotz_1995, Saad_2003, Kalita_2014}. After solving equations (\ref{Main_governing_eq_8}) - (\ref{Main_governing_eq_10}) with a constant initial pressure, the subsequent step involves determining the pressure $p$ for the next time step. Determining the pressure is a significant challenge when using the primitive variable approach to solve the Navier-Stokes equations, as there is no explicit pressure term.
We opted to utilize the modified compressibility technique proposed by Cortes and Miller \cite{Cortes_1994} for solving the pressure problem. This approach was chosen for its efficiency, simplicity, and straightforward implementation.
In this approach, the modified continuity equation is expressed as:
$$
\lambda \nabla \cdot \mathbf{v}+p=0 .
$$
At each time step, after computing pressure gradients and solving momentum equations, we compute the dilation parameter $D = v_y + u_x + w_z$. If the greatest absolute value of $D$, indicated as $|D|_{\max}$, falls below a predetermined tolerance threshold, we conclude that the pressure value has met the required level of accuracy and we can proceed to the next time level. If the maximum value of $|D|$ exceeds the specified tolerance threshold, we initiate a pressure correction step to improve the accuracy of the pressure value:
$$
p^{n1+1}=p^{n1}-\lambda \nabla \cdot \mathbf{v} .
$$
Here, $p^{n1+1}$ denotes the updated pressure, $p^{n1}$ represents the obtained pressure value in the preceding pressure iteration, and $\lambda$ signifies a relaxation parameter. This iterative procedure continues until the maximum absolute value of $|D|_{\max}$ satisfies the specified tolerance limit. Afterward, one can proceed to the next time step, and the process repeats until a steady state is achieved.\\
\section{Sensitivity Test and Scheme Validation}
\label{sec:SENSITIVITY TESTS AND SCHEME VALIDATION}
\subsection{Grid and Time Independence Test}
To validate the grid independence of our results and optimize computational efficiency, we conducted a study to assess the effect of varying mesh size while keeping other parameters constant. We examined four distinct grid dimensions: $21\times21\times21$, $41\times41\times41$, $81\times81\times81$, and $161\times161\times 161$. Meanwhile, we kept the values of $\Delta \tau$, $Re$ , and $n$ constant at $0.02, 100,$ and $0.5$, respectively.
Table \ref{grid_independent_test} illustrates the velocity values for each grid size at two designated monitoring points $(0.65, 0.65, 0.65)$ and $(0.45, 0.45, 0.45)$, situated near the core region of the cavity at a dimensionless time, $\tau=100$. Our investigation indicates that the maximum relative error between the computed values on $81\times81\times81$ and $161\times161\times161$ grids is merely $1.52\%$. This result suggests that an increase in the grid size has no significant effect on the computed results. Therefore, the mesh size of $81\times81\times81$ is considered for the present computational investigation. In the time sensitivity test, as depicted in Table \ref{time_independent_test}, it is apparent that the maximum variation in the simulated results occurs when comparing $\Delta \tau = 0.002$ and $\Delta \tau = 0.02$, with only a maximum relative error ($\delta_e$(\%)) of 0.78\%. This observation suggests that $\Delta \tau = 0.02$ is enough for accurately capturing flow phenomena. Hence, in our endeavor to optimize both computational efficiency and solution accuracy, we opt for a grid size of $81\times81\times81$ and $\Delta \tau = 0.02$ for our computational study.

{\small\begin{table}[htbp]
\caption{\small Velocity values at two designated monitoring points ($0.65, 0.65, 0.65$) and ($0.45, 0.45, 0.45$), near the core region of the cavity with fixed $Re = 100$, $n=0.5$, $\tau=100$ and  $\Delta \tau$ = 0.02 by employing four distinct sizes of grid}\label{grid_independent_test}
\centering
 \begin{tabular}{cccccc}  \hline \hline
Monitoring point & M$\times$N$\times$O    &    $u$   &  $v$  &  $w$ & $\delta_e$(\%)    \\ \hline
\hline
(0.65, 0.65, 0.65) & (21 $\times$ 21 $\times$ 21) &  -0.09312 &   -0.00913    &   -0.01217  & --- \\
& (41 $\times$ 41 $\times$ 41)                     &  -0.11761  &  -0.01085  &   -0.00876 & 28.0 \\
& (81 $\times$ 81 $\times$ 81)                     &  -0.13480 &  -0.01127  &   -0.00722  & 17.5  \\
& (161 $\times$ 161 $\times$ 161)              &  -0.13403 &  -0.01134  &   -0.00714  & 1.10 \\
\hline
(0.45, 0.45, 0.45) & (21 $\times$ 21 $\times$ 21)   &  -0.07117 &   -0.00102    &   0.01797 & --- \\
& (41 $\times$ 41 $\times$ 41)                      &  -0.08746   &  -0.00119  &   0.02361 & 31.38 \\
& (81 $\times$ 81 $\times$ 81)                       &  -0.09964 &  -0.00131  &   0.02791  & 18.21 \\
& (161 $\times$ 161 $\times$ 161)                    &  -0.10116 &  -0.00130  &   0.02802  & 1.52  \\
\hline
\hline
 \end{tabular}
\end{table}
}

{\small\begin{table}[htbp]
\caption{\small Velocity values at two specified monitoring points ($(0.65, 0.65, 0.65)$ and $(0.45, 0.45, 0.45)$) near the core region of the cavity with time $=100$, $n = 0.5$, $Re = 100$ and 81 $\times$ 81 $\times$ 81 grid resolution by employing three distinct time step}\label{time_independent_test}
\centering
\begin{tabular}{cccccc}  \hline \hline
Monitoring point & $\Delta \tau$   &    $u$   &  $v$  &  $w$ & $\delta_e$(\%)     \\ \hline
(0.65, 0.65, 0.65) & 0.2     &  -0.14023 &   -0.01146    &   -0.00699  & --- \\
&0.02                        & -0.13480 &  -0.01127  &   -0.00722 & 3.87\\
&0.002                       &  -0.13374 &  -0.01119  &   -0.00727  & 0.78  \\ \hline
(0.45, 0.45, 0.45) & 0.02  &  -0.09637 &  -0.00127    &   0.02695  & -- \\
&0.02                       &  -0.09964 &  -0.00131  &   0.02791 & 3.56\\
&0.002                      &  -0.09943 &  -0.00130  &   0.02783  & 0.28 \\
\hline
 \end{tabular}
\end{table}
}

\subsection{Validation of the Proposed Scheme for non-Newtonian Power-Law Model}
\begin{figure}
    \centering
    \includegraphics[width=\textwidth]{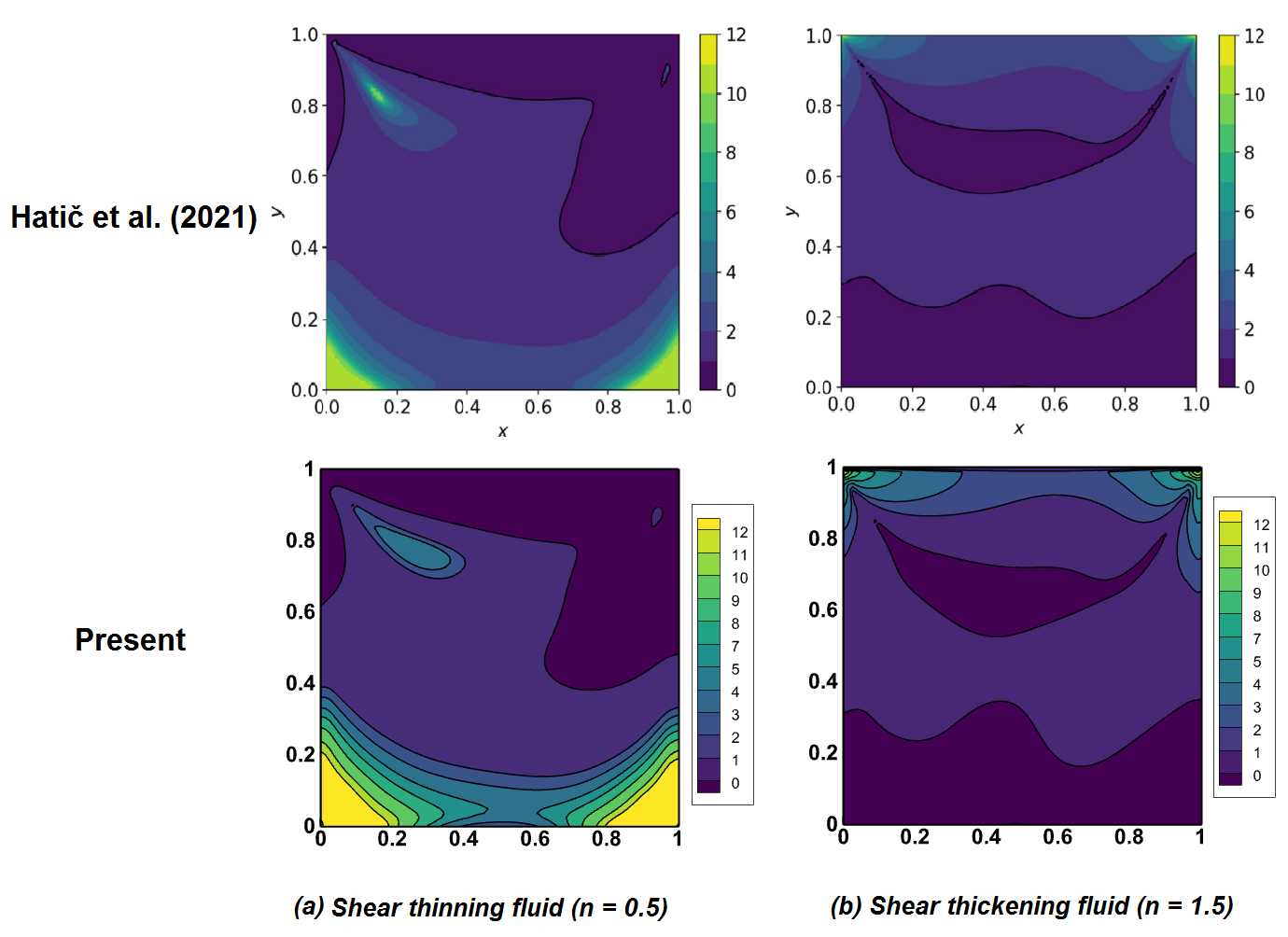}
    \caption{Comparison of the viscosity contours between present results and the results of Hatič et al. (2D) \cite{Hatič_2021} for shear-thinning ($n=0.5)$ and shear-thickening $(n=1.5)$ non-Newtonian fluids}
    \label{fig:viscosity_comparison}
\end{figure}

\begin{figure}
    \centering
    \includegraphics[width=\textwidth]{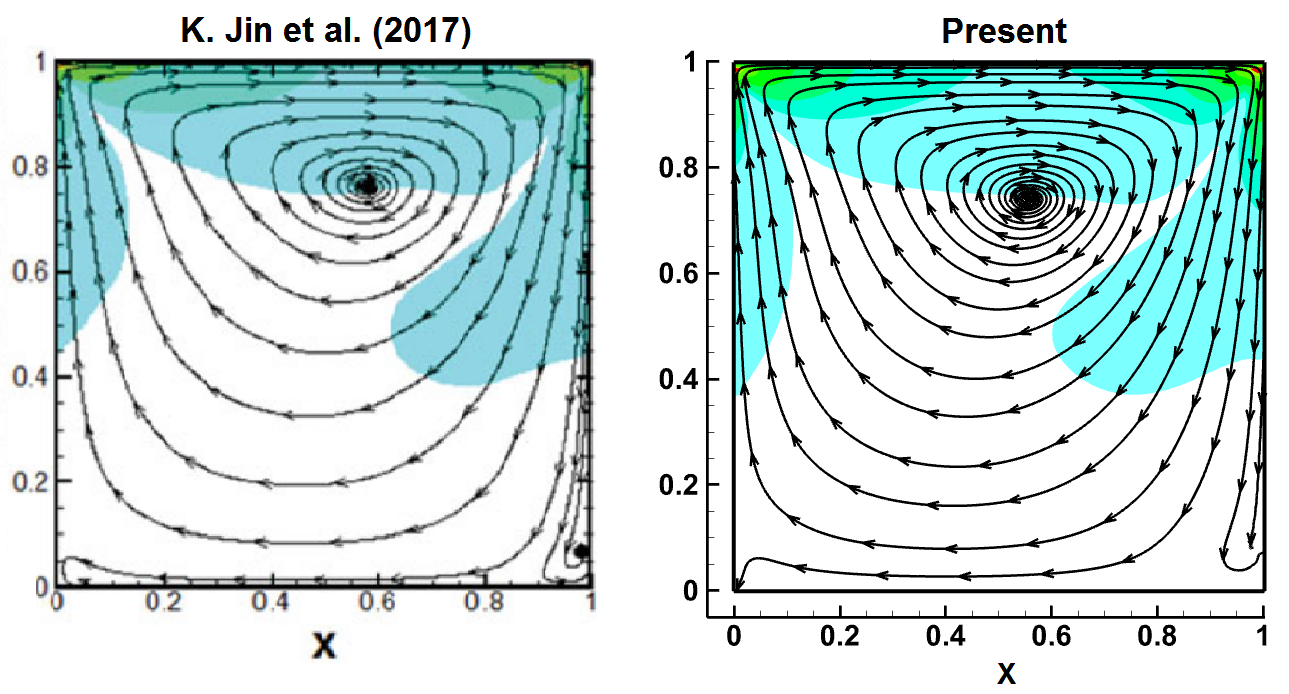}
    \caption{Comparison of the streamlines pattern between present result and benchmark result of Jin et al. (3D) \cite{Jin_2017} at $Re=100$ and $n=1.5$}
    \label{fig:streamlines_comparison}
\end{figure}

\begin{figure}
    \centering
    \includegraphics[width=\textwidth]{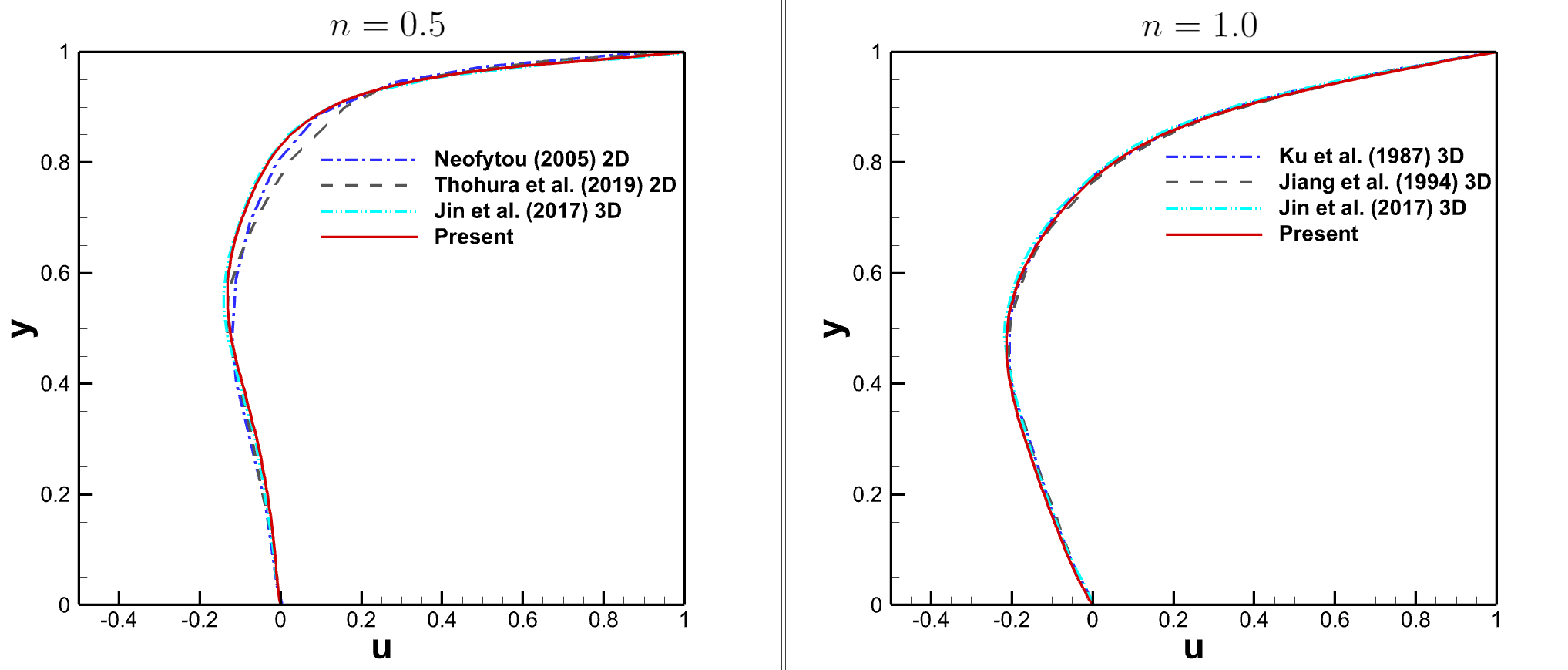}
    \caption{ Comparison of $u$-velocity profiles between present results with some
published results \cite{Neofytou_2005,Jin_2017,Thohura_2019} for shear-thinning ($n=0.5$) and Newtonian ($n=1.0$) fluids at $Re=100$}
    \label{fig:centerline_comparison}
\end{figure}
To validate the robustness and credibility of the proposed HOSC scheme, we performed a thorough validation against previously published results for the power-law fluid model in a closed cavity. Moreover, it should be noted here that experimental data for non-Newtonian power-law fluid flow in cubical cavities have not been reported. Therefore, we opted to utilize benchmark computational results from the existing literature that are widely recognized and accepted within the field of non-Newtonian fluids. We examine a range of power-law index ($n$) values to assess the performance of the proposed scheme across various flow regimes. Validation is conducted both quantitatively and qualitatively by comparing our results with those reported in the literature.
In Figures \ref{fig:viscosity_comparison} and \ref{fig:streamlines_comparison}, we offer visual comparisons of viscosity and streamlines, respectively, between our findings and those documented in the literature \cite{Jin_2017, Hatič_2021} for both shear-thinning $(n=0.5)$ and shear-thickening $(n=1.5)$ fluids. Figure \ref{fig:centerline_comparison} illustrates the comparison of the centerline-velocity graph $u$ with references \cite{Neofytou_2005, Jin_2017, Thohura_2019} for $n=0.5$ and $n=1.0$. Table \ref{primary_vortex_comparison} presents a quantitative comparison of the primary vortex location at $Re=100$ for different $n$ values $(= 0.5, 1.0, 1.5)$ with the benchmark results of \cite{Jin_2017}. We found that results obtained with our proposed scheme agree well with previous published 3D work which again confirms the accuracy of the developed code and scheme.\\
It is noteworthy that our results demonstrate an excellent agreement with previously published findings, demonstrating the high degree of consistency, precision, and dependability of our technique in capturing the complicated 3D phenomena under the non-Newtonian power-law model.

{\small\begin{table}[htbp]
\caption{\small  Comparison of location of primary vortex core (at $z=0.5$ plane) for Newtonian and non-Newtonian fluid flows at $Re=100$}\label{primary_vortex_comparison}
\centering
 \begin{tabular}{cccc}  \hline \hline
$n$   &  &    Jin et al. 3D (2017) \cite{Jin_2017}   &  Present    \\ \hline
\hline
 0.5 &$x_c $ &  0.716 &   0.716     \\
 &$y_c$  &  0.815 &   0.818     \\ \hline
 1.0 &$x_c$  &  0.619 &   0.619    \\
 &$y_c $ &  0.762 &   0.756     \\ \hline
 1.5 &$x_c$  &  0.579 &   0.569     \\
 &$y_c$  &  0.763 &   0.745     \\
\hline
\hline
 \end{tabular}
\end{table}
}

\section{Results and Discussion}
\label{sec:Results and Discussion}
This section unveils the outcomes derived from numerical simulations, focusing on the influence of the Reynolds number ($Re$) and power-law index ($n$) on the flow regime, velocity, viscosity, and pressure distribution within the 3D cubic cavity. The findings regarding flow in a 3D lid-driven cubic cavity for a non-Newtonian power-law fluid using the higher-order finite difference method are unprecedented in the literature. Consequently, these results can serve as benchmark solutions for other researchers. Here, we explore three distinct power-law indices: $n = 0.5$, $n = 1.0$, and $n = 1.5$, across four different Reynolds numbers: $Re = 1$, $Re = 50$, $Re =100$, and $Re =200$. All simulations are conducted on a uniform grid size of $81 \times 81 \times 81$. \\

\begin{figure}[htbp]
 \centering
 \vspace*{0pt}%
 \hspace*{\fill}%
\begin{subfigure}{1.0\textwidth}     
    \centering
    \includegraphics[width=\textwidth]{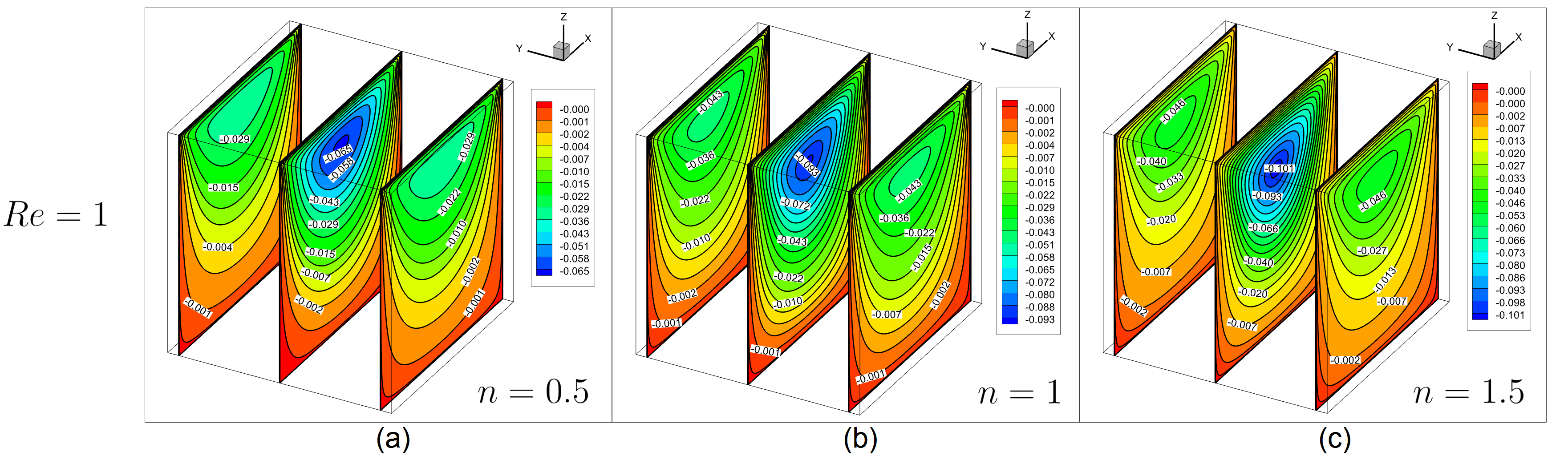}%
    \captionsetup{skip=2pt}%
  \end{subfigure}%
  \hspace*{\fill}

  \vspace*{8pt}%
  \hspace*{\fill}%
  \begin{subfigure}{1.0\textwidth}     
    \centering
    \includegraphics[width=\textwidth]{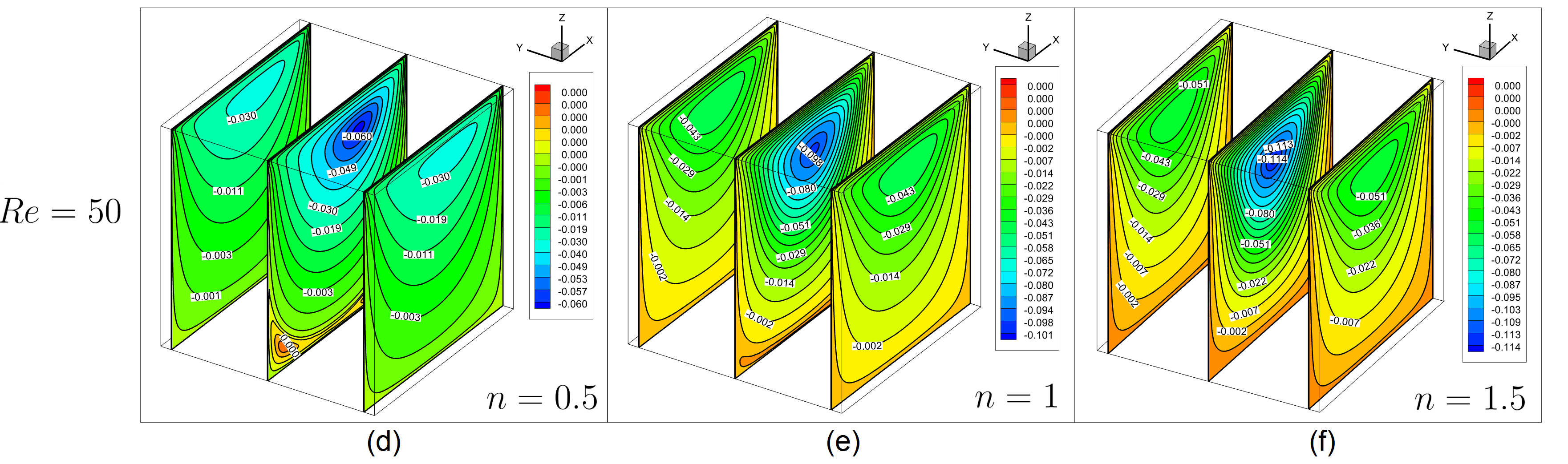}%
    \captionsetup{skip=2pt}%
  \end{subfigure}%

  \vspace*{8pt}%
  \hspace*{\fill}%
  \begin{subfigure}{1.0\textwidth}     
    \centering
    \includegraphics[width=\textwidth]{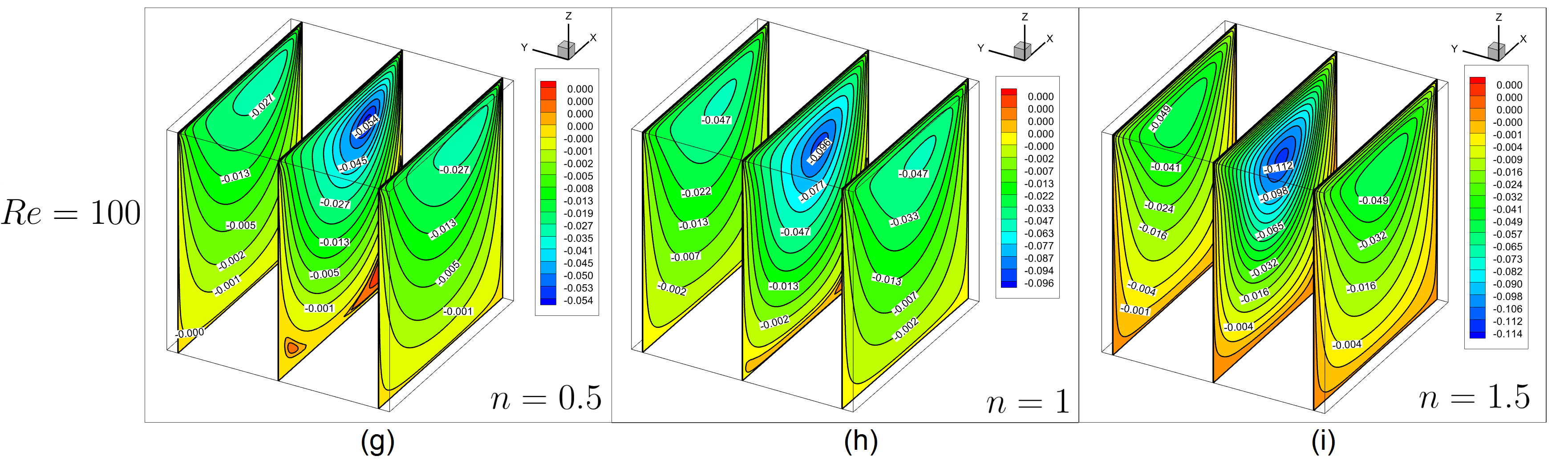}%
    \captionsetup{skip=2pt}%
  \end{subfigure}%
  \hspace*{\fill}

  \vspace*{8pt}%
  \hspace*{\fill}%
  \begin{subfigure}{1.0\textwidth}     
    \centering
    \includegraphics[width=\textwidth]{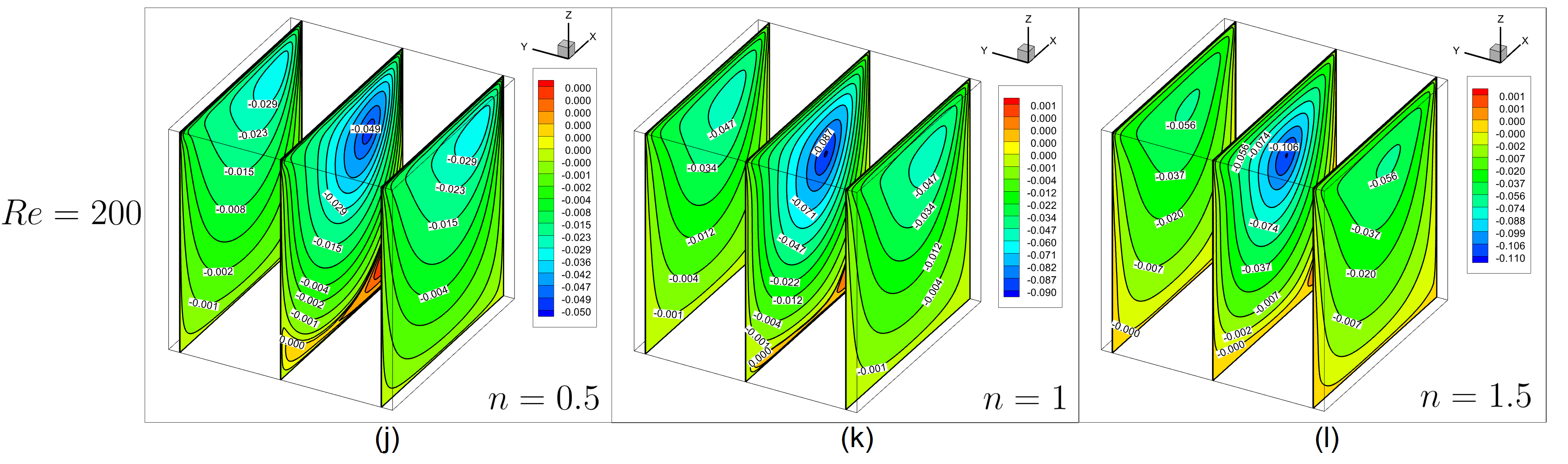}%
    \captionsetup{skip=2pt}%
  \end{subfigure}%
  \hspace*{\fill}
  \vspace*{2pt}%
  \hspace*{\fill}%
  \caption{Streamline visualization on the three slices ($y = 0.05$, $y = 0.5$ and $y = 0.95$ planes) for the lid-driven cubic cavity problem}
  \label{fig:Streamlines_3_plane}
\end{figure}

\begin{figure}[htbp]
 \centering
 \vspace*{0pt}%
 \hspace*{\fill}%
\begin{subfigure}{1.0\textwidth}     
    \centering
    \includegraphics[width=\textwidth]{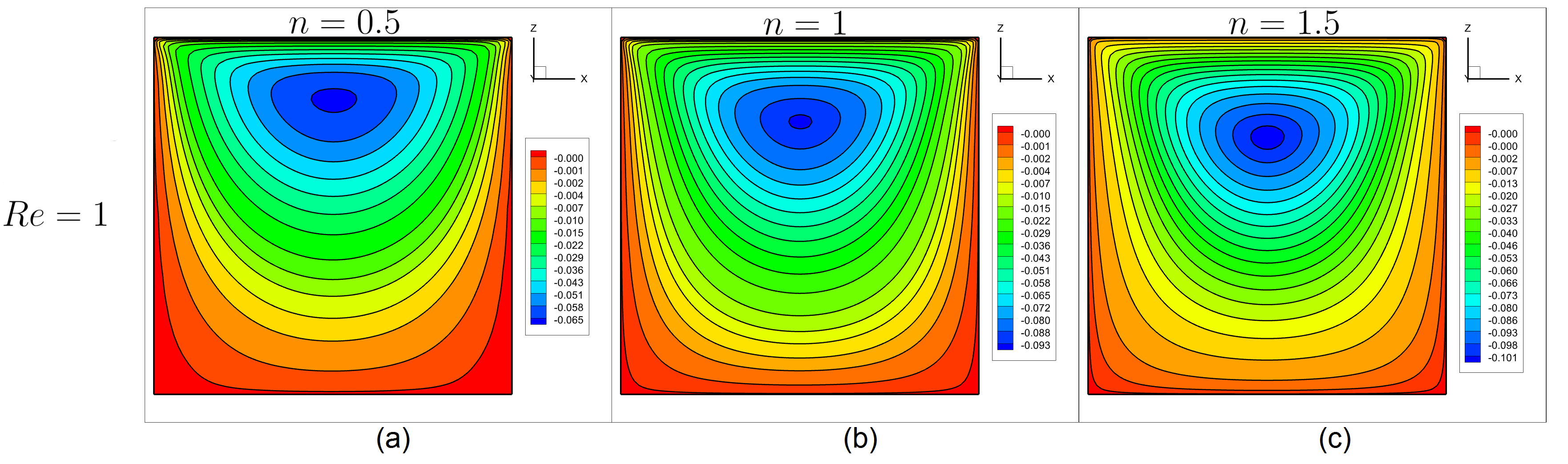}%
    \captionsetup{skip=2pt}%
  \end{subfigure}%
  \hspace*{\fill}

  \vspace*{8pt}%
  \hspace*{\fill}%
  \begin{subfigure}{1.0\textwidth}     
    \centering
    \includegraphics[width=\textwidth]{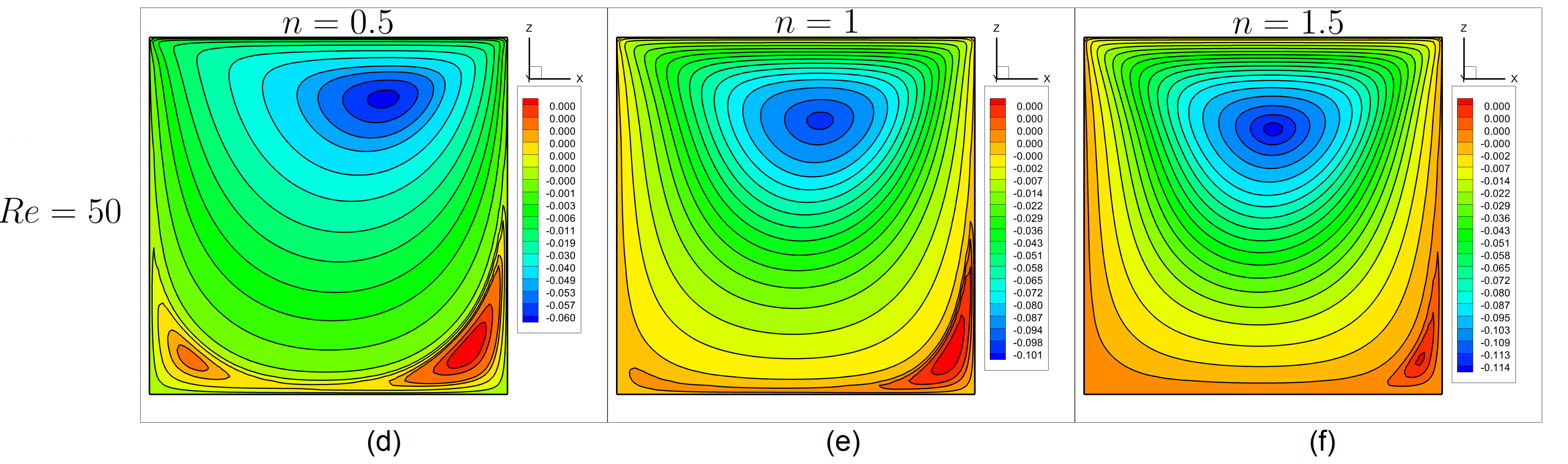}%
    \captionsetup{skip=2pt}%
  \end{subfigure}%

  \vspace*{8pt}%
  \hspace*{\fill}%
  \begin{subfigure}{1.0\textwidth}     
    \centering
    \includegraphics[width=\textwidth]{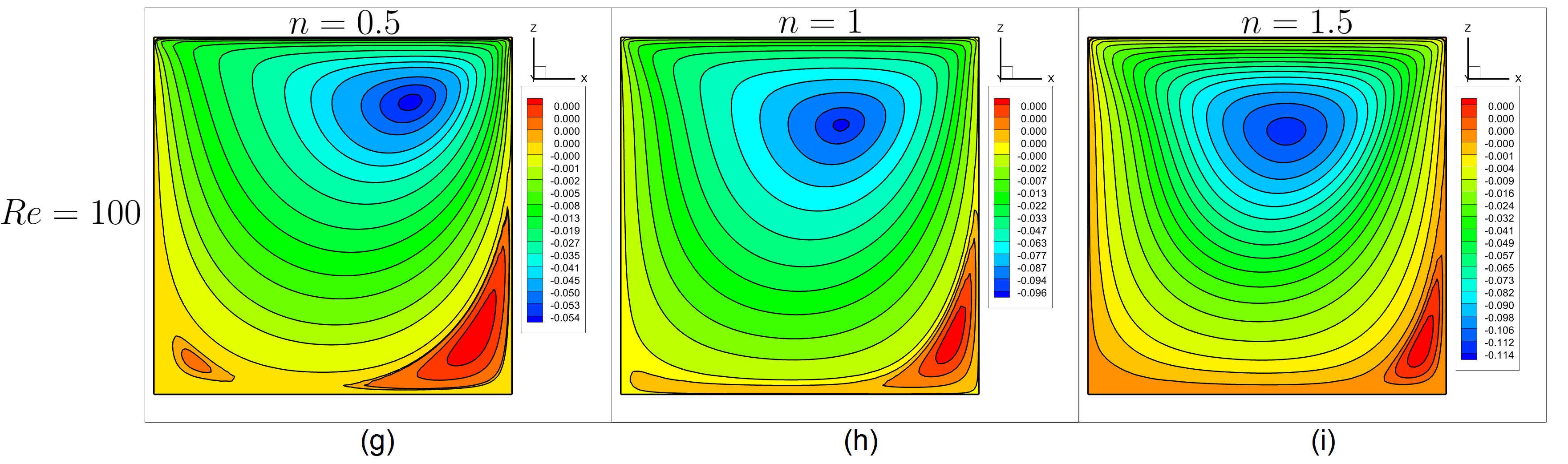}%
    \captionsetup{skip=2pt}%
  \end{subfigure}%
  \hspace*{\fill}

  \vspace*{8pt}%
  \hspace*{\fill}%
  \begin{subfigure}{1.0\textwidth}     
    \centering
    \includegraphics[width=\textwidth]{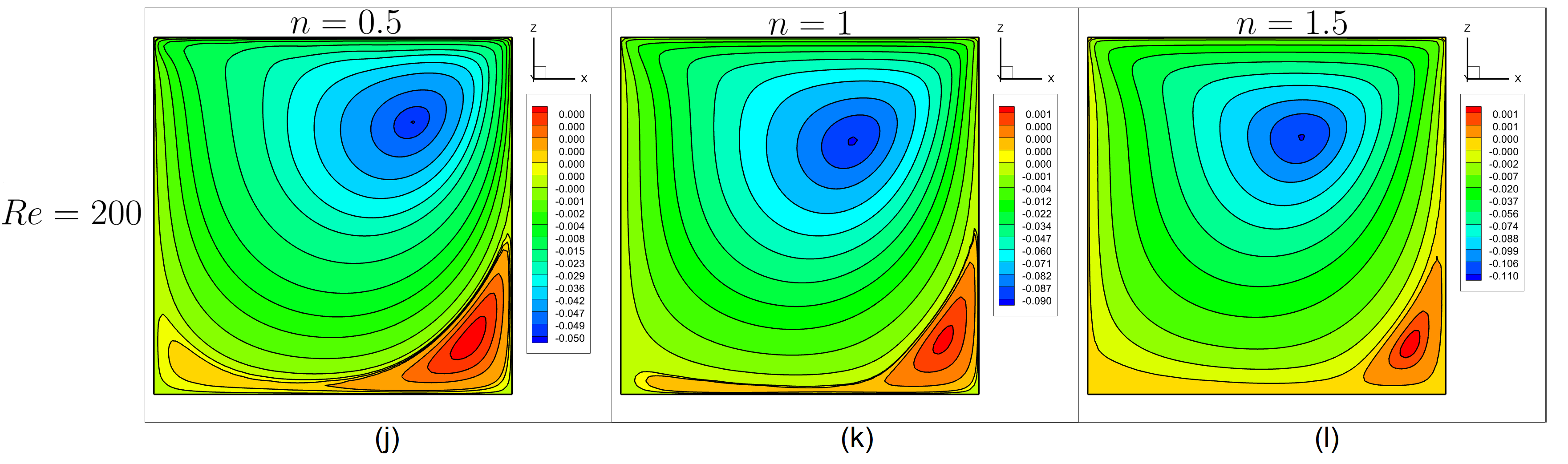}%
    \captionsetup{skip=2pt}%
  \end{subfigure}%
  \hspace*{\fill}
  \vspace*{2pt}%
  \hspace*{\fill}%
  \caption{Streamline visualization on the middle of y-plane ($y = 0.5$) for the lid-driven cubic cavity problem}
  \label{fig:Streamlines_mid_plane_plane}
\end{figure}

{\small\begin{table}[htbp]
\caption{\small Location $(x, z)$, of the primary vortex core at the middle of the $y$ plane for different $Re$ and $n$ values}\label{primary_vortex}
\centering
 \begin{tabular}{cccc}  \hline \hline
$Re$     &    $n=0.5$   &  $n=1.0$  &  $n=1.5$   \\ \hline
\hline
1  &  (0.503, 0.825) &   (0.501, 0.764)    &  (0.501, 0.720)   \\
50  &  (0.655, 0.826) &  (0.569, 0.766)  &   (0.528, 0.743)    \\
100   &  (0.716, 0.818) &  (0.619, 0.756)  &   (0.569, 0.745)   \\
200     &  (0.724, 0.760) &  (0.649, 0.709)  &   (0.598, 0.718)   \\
\hline
\hline
 \end{tabular}
\end{table}
}
In Figure \ref{fig:Streamlines_3_plane}, we depict the streamlines contours across three distinct planes ($y=0.05$, $y=0.5$, $y=0.95$), showcasing the dynamic interplay of Reynolds number ($Re$) and power-law index ($n$). For additional visualization at the symmetric plane ($y=0.5$), the streamlines pattern at the symmetric plane ($y=0.5$) is highlighted in Figure \ref{fig:Streamlines_mid_plane_plane} for various $Re$ and $n$ values. At $Re=1$, our observations (Figure \ref{fig:Streamlines_3_plane} (a,b,c)) indicate the presence of a single large primary vortex on each considered plane. The presence of negative stream function values signifies a clockwise rotation of the fluid, consistently observed across all considered values of $n$. The location $(x,z)$ of the primary vortex at $y=0.5$ plane is presented in Table \ref{primary_vortex}. Notably, as $n$ increases, the primary vortex shifts downwards, which can also be visualized from Figure \ref{fig:Streamlines_mid_plane_plane}. For $Re=50$, in the shear-thinning ($n=0.5$) scenario, the emergence of two secondary vortices (Figure \ref{fig:Streamlines_mid_plane_plane} (d)) at the bottom of the cavity corners is observable on the $y=0.5$ plane. However, this phenomenon diminishes in prominence with a shift towards shear thickening $(n=1.5)$. Additionally, the size of the secondary vortex existing on the left bottom is relatively small to the vortex existing on the right bottom side of the cavity. A discernible trend is the gradual migration of the primary vortex towards the center of the cavity, as $n$ increases. The absence of a secondary vortex is evident on the other two planes ($y=0.05$ and $y=0.95$) for all $n$ and $Re$. Again, for high Reynolds numbers ($Re=100,200$), we observe a shift in the center of the primary vortex towards the center of the cavity, when transitioning from a shear-thinning $(n=0.5)$ to a shear-thickening fluid $(n=1.5)$. Hence, we can conclude that variations in the power-law index $(n)$ have a significant impact on the fluid flow within the cavity. When examining the effect of Reynolds number $(Re)$, for the shear-thinning fluid $(n=0.5)$, two secondary corner vortices are observed near the bottom wall of the cavity for $Re \geq 50$. However, the sizes of these vortices vary for different $Re$. Specifically, the left vortex diminishes while the right vortex expands with increasing $Re$. It is also interesting to note that there is no secondary vortex formulation near the left bottom corner for the case of shear thickening $(n=1.5)$. The center of the primary vortex is shifting toward the right top side of the cavity as we increase the $Re$ at any specific $n$.

A correlation between $Re$, $n$, and the location of the primary vortex is noticeable. Specifically, the center of the primary vortex tends to shift towards the cavity's center with an increase of $n$ values at a given $Re$. Additionally, as $Re$ increases at a specific $n$, the center of the primary vortex tilts towards the right top corner of the cavity. However, this displacement is more pronounced for shear-thinning fluids ($n=0.5$). This trend is further supported by the data presented in Table \ref{primary_vortex}. Similar flow patterns were observed by other researchers as well \cite{Kalita_2014, Jin_2017, Jabbari_2019, Hatič_2021}. \\

\begin{figure}
    \centering
    \includegraphics[width=\textwidth]{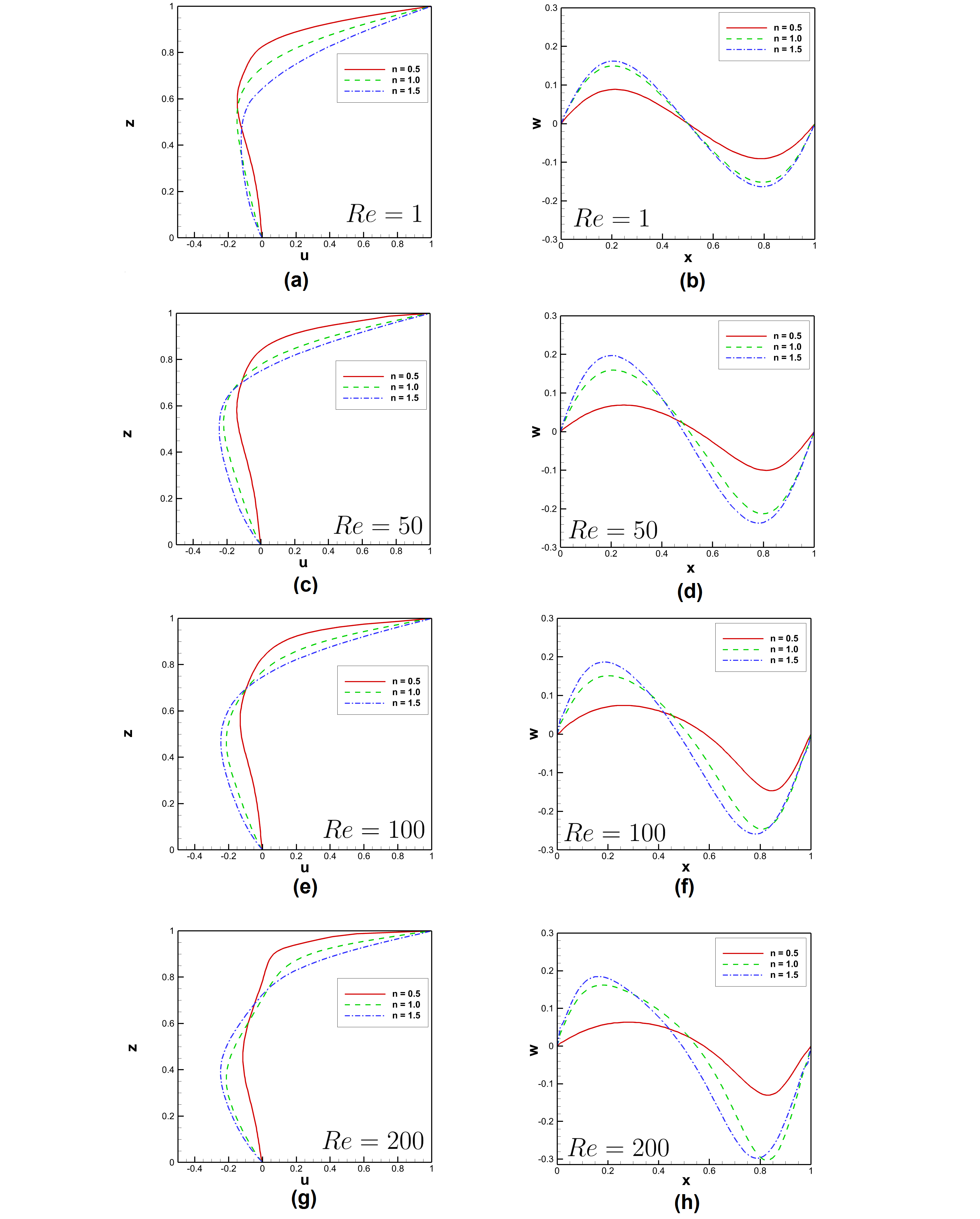}
    \caption{Centerline velocities variation for different $Re$ and $n$ values}
    \label{fig:Velocity_Variation}
\end{figure}

Figure \ref{fig:Velocity_Variation} shows the axial velocity profiles for $u$ vs $z$ at $x = 0.5$ and $w$ vs $x$ at $z = 0.5$ for different $Re$ and $n$ values. In the velocity profile depicted in Figure \ref{fig:Velocity_Variation} (a), it's noted that the $u$ component velocity near the bottom of the cavity is negative, transitioning to positive as it nears the top lid $(z=1)$ for $Re=1$ across all $n$ values. Similarly, due to the rotating vortices within the cavity, the $w$ component velocity near the left wall $(x=0.0)$ is positive, whereas it tends to be negative towards the right wall (Figure \ref{fig:Velocity_Variation}(b)). This observation confirms the clockwise rotation of the fluid within the cavity. A similar pattern emerges for different $Re$ values, but the magnitude varies depending on the intensity of the vortices within the cavity. In the case of shear-thinning fluid $(n=0.5)$, the $u$ velocity near the top moving wall surpasses that of Newtonian $(n=1.0)$ and shear-thickening fluid $(n=1.5)$ for all $Re$. Conversely, a reverse trend in $u$ velocity is observed near the bottom of the cavity. This phenomenon arises from the viscosity decrease with increasing shear rate in shear-thinning fluids. Consequently, the reduced viscosity leads to diminished resistance to flow under a high-shear rate, leading to increased velocity near the lid. Across all $n$ values, increasing $Re$ induces higher gradients alongside increased peak values for $u$ and $w$. This phenomenon is attributed to the gradual reduction of viscous effects with rising $Re$. Additionally, it's noticeable that with higher $n$ values, the peak magnitude of $u$ velocity decreases, while the peak magnitude of $w$ velocity increases for all $Re$ values. Thus, it's apparent that both the power-law index $(n)$ and Reynolds number $(Re)$ significantly impact the velocity profile within the cavity.\\

\begin{figure}[htbp]
 \centering
 \vspace*{0pt}%
 \hspace*{\fill}%
\begin{subfigure}{1.0\textwidth}     
    \centering
    \includegraphics[width=\textwidth]{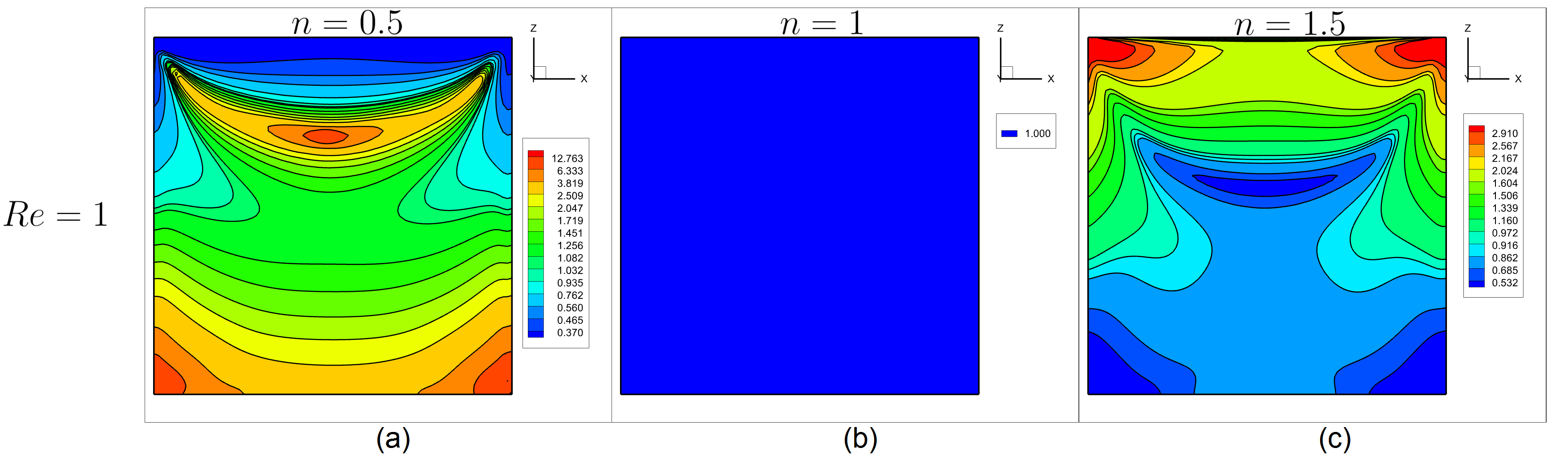}%
    \captionsetup{skip=2pt}%
  \end{subfigure}%
  \hspace*{\fill}

  \vspace*{8pt}%
  \hspace*{\fill}%
  \begin{subfigure}{1.0\textwidth}     
    \centering
    \includegraphics[width=\textwidth]{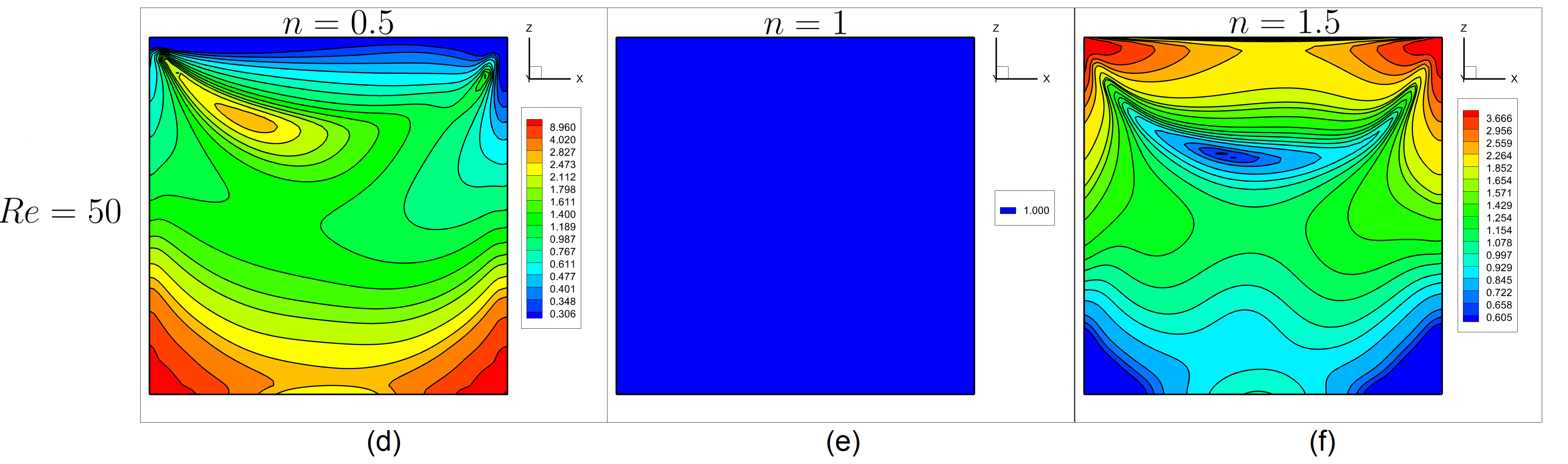}%
    \captionsetup{skip=2pt}%
  \end{subfigure}%

  \vspace*{8pt}%
  \hspace*{\fill}%
  \begin{subfigure}{1.0\textwidth}     
    \centering
    \includegraphics[width=\textwidth]{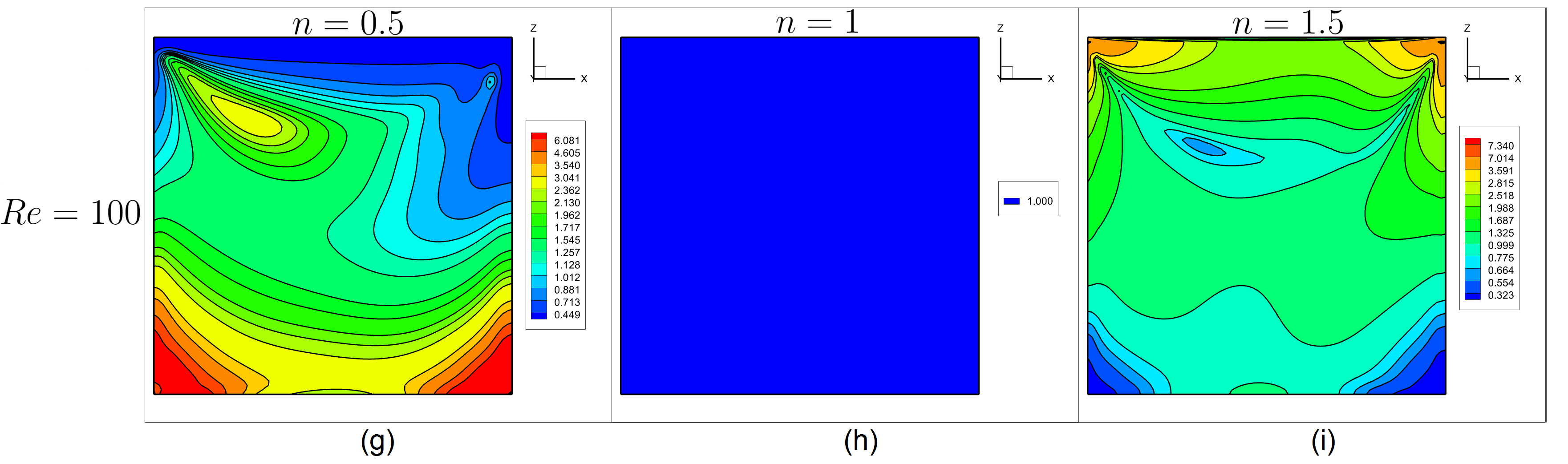}%
    \captionsetup{skip=2pt}%
  \end{subfigure}%
  \hspace*{\fill}

  \vspace*{8pt}%
  \hspace*{\fill}%
  \begin{subfigure}{1.0\textwidth}     
    \centering
    \includegraphics[width=\textwidth]{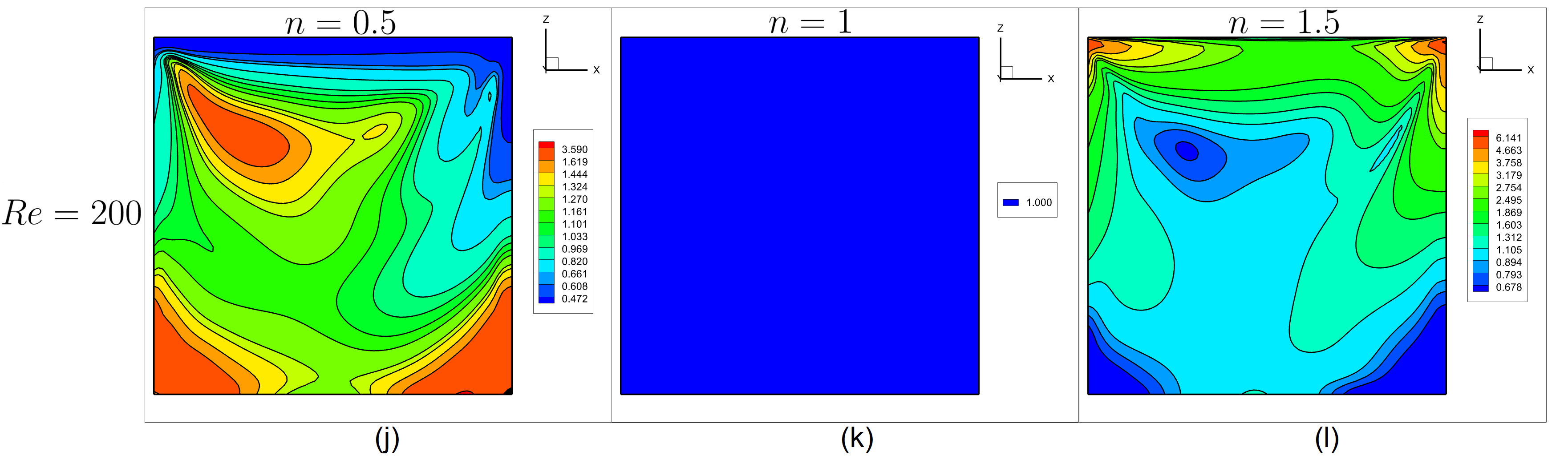}%
    \captionsetup{skip=2pt}%
  \end{subfigure}%
  \hspace*{\fill}
  \vspace*{2pt}%
  \hspace*{\fill}%
  \caption{Viscosity contours on the middle of y-plane ($y = 0.5$) for different $Re$ and $n$ values}
  \label{fig:Viscosity_mid_plane_plane}
\end{figure}

In Figure \ref{fig:Viscosity_mid_plane_plane}, we present the viscosity contours for different $Re$ and $n$ values at the $y=0.5$ plane. The impact of the $n$ and $Re$ can be observed in row-wise and column-wise sub-figures of Figure \ref{fig:Viscosity_mid_plane_plane}, respectively. We observed that the viscosity is low near the top moving lid for shear-thinning fluids and high near the moving lid for shear-thickening fluids. This discrepancy arises from the behavior of these fluids: shear-thinning fluids exhibit viscosity reduction as the shear rate increases, and since the fluid experiences higher shear rates near the moving lid, viscosity decreases accordingly. Conversely, shear-thickening fluids demonstrate the opposite trend, with viscosity increasing as the shear rate rises. The viscosity remains unchanged for Newtonian fluids. In all scenarios involving shear-thinning fluids, regions of higher viscosity are noticeable at the bottom corners of the cavity, whereas in shear-thickening cases, heightened viscosity is observed near the top corners. This distinction highlights the significant impact of $n$ on viscosity contours. Additionally, at $Re=1$, the viscosity contours exhibit symmetry at $x=0.5$. However, asymmetry emerges as $Re$ increases for both shear-thinning and shear-thickening fluids. Particularly intriguing is the behavior observed in shear-thinning fluids, where the region of low viscosity expands with increasing Reynolds number ($Re$), given that viscosity decreases with higher shear rates. This phenomenon can be attributed to the interaction between shear-thinning behavior and the flow dynamics associated with higher $Re$. Conversely, in the case of shear-thickening fluids, the region of high viscosity expands with increasing $Re$. The reason for this behavior is that as $Re$ increases, so do the flow velocities and shear rates near the sliding lid. Similar viscosity contours were observed in the work of Hatič et al. \cite{Hatič_2021}.\\

\begin{figure}[htbp]
 \centering
 \vspace*{0pt}%
 \hspace*{\fill}%
\begin{subfigure}{1.0\textwidth}     
    \centering
    \includegraphics[width=\textwidth]{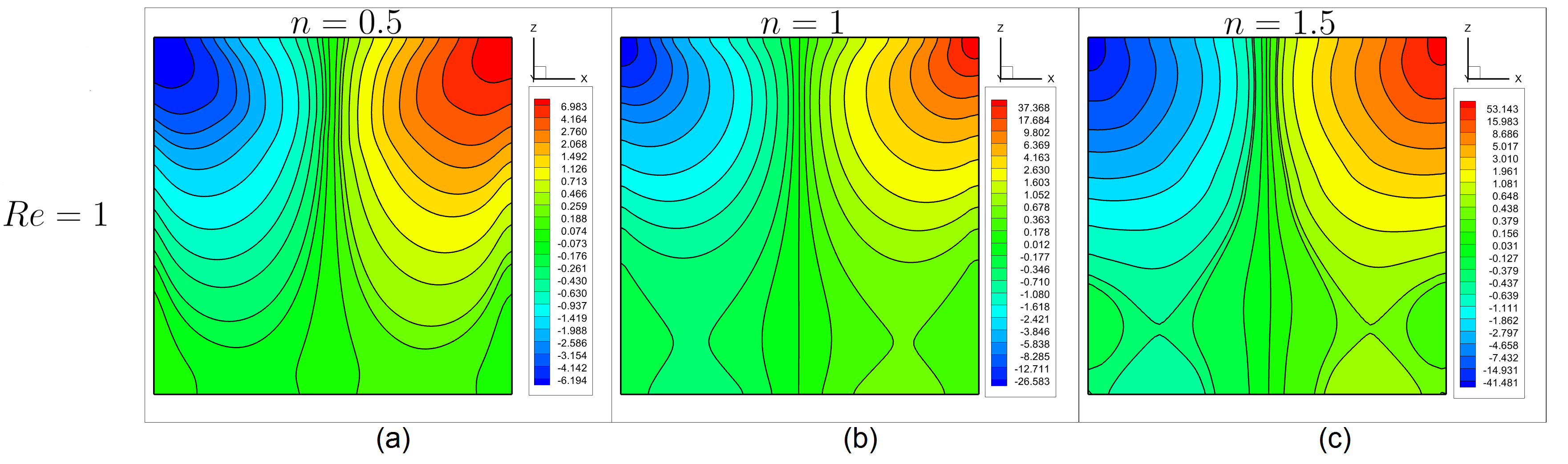}%
    \captionsetup{skip=2pt}%
  \end{subfigure}%
  \hspace*{\fill}

  \vspace*{8pt}%
  \hspace*{\fill}%
  \begin{subfigure}{1.0\textwidth}     
    \centering
    \includegraphics[width=\textwidth]{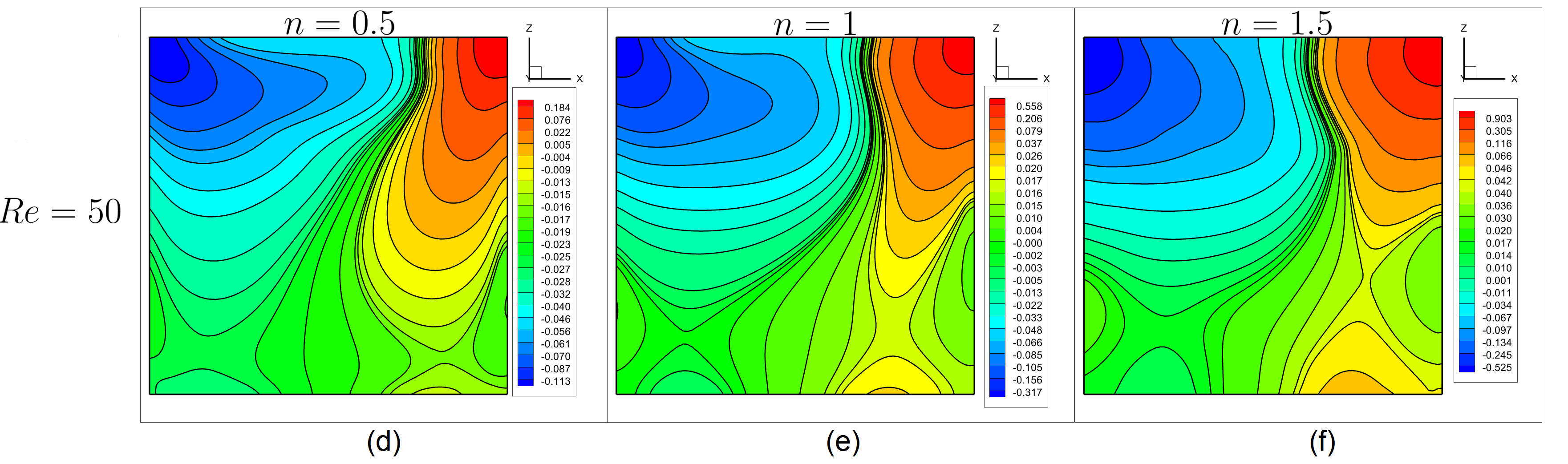}%
    \captionsetup{skip=2pt}%
  \end{subfigure}%

  \vspace*{8pt}%
  \hspace*{\fill}%
  \begin{subfigure}{1.0\textwidth}     
    \centering
    \includegraphics[width=\textwidth]{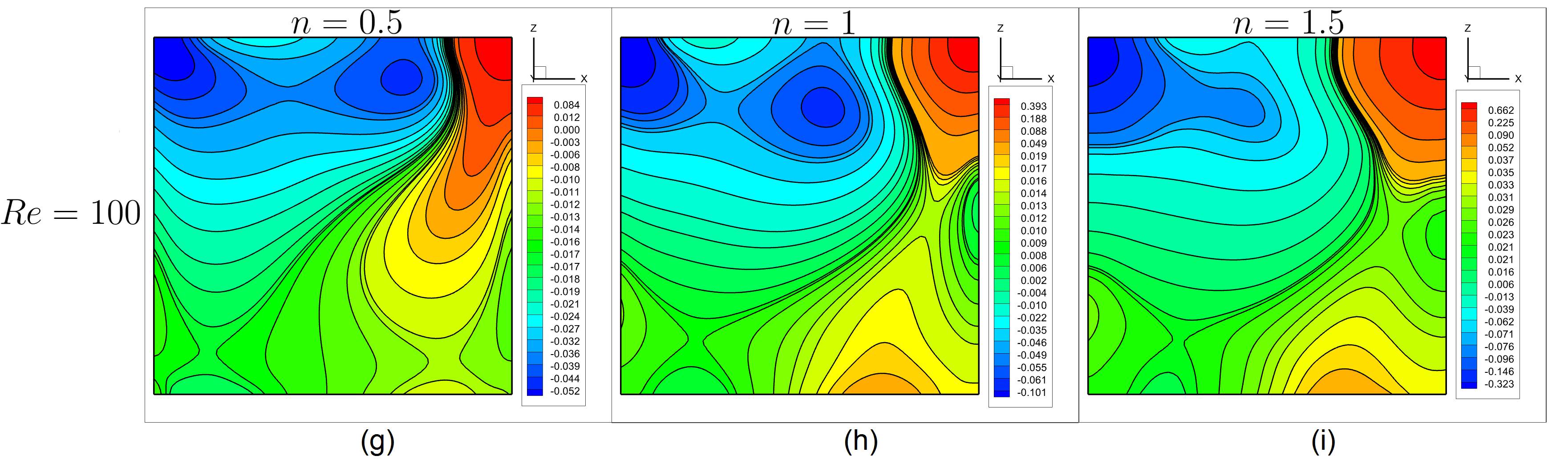}%
    \captionsetup{skip=2pt}%
  \end{subfigure}%
  \hspace*{\fill}

  \vspace*{8pt}%
  \hspace*{\fill}%
  \begin{subfigure}{1.0\textwidth}     
    \centering
    \includegraphics[width=\textwidth]{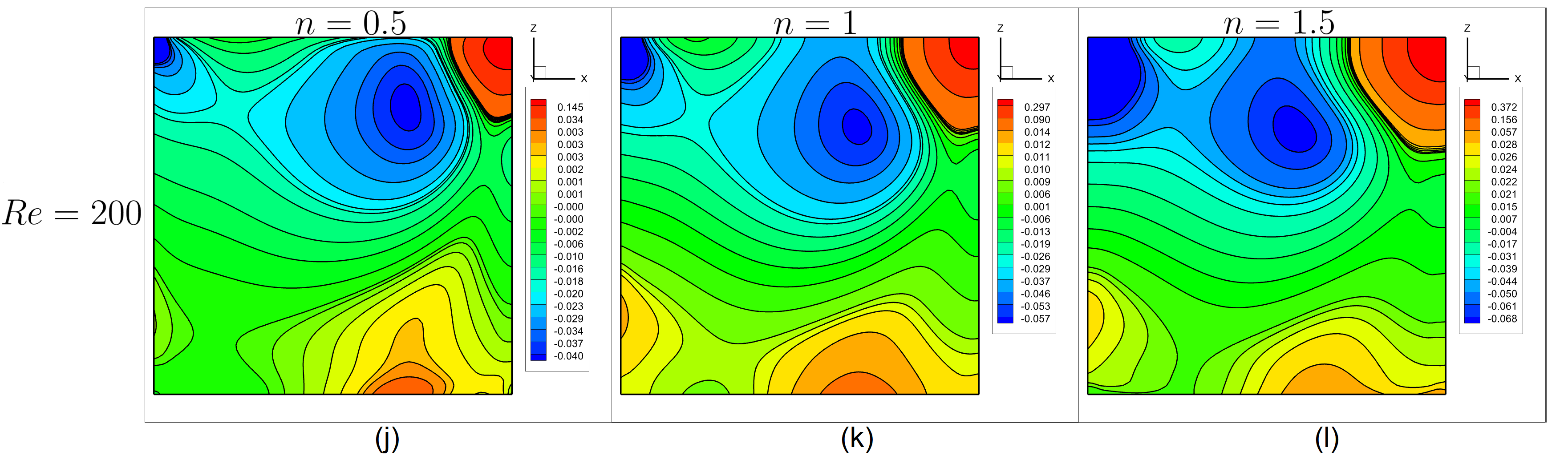}%
    \captionsetup{skip=2pt}%
  \end{subfigure}%
  \hspace*{\fill}
  \vspace*{2pt}%
  \hspace*{\fill}%
  \caption{Pressure distribution on the middle of y-plane ($y = 0.5$) for different $Re$ and $n$ values}
  \label{fig:Pressure_mid_plane_plane}
\end{figure}

It is worth mentioning that there are very few published papers that exhibit pressure contours. Many studies use streamfunction-vorticity formulations of Navier-Stokes equations in 2D, which often eliminate pressure. However, it's crucial to emphasize that the pressure field holds significant importance, offering invaluable insights into fluid behavior. Figure \ref{fig:Pressure_mid_plane_plane} display the pressure contours at the middle of the $y$-plane ($y=0.5$) for various $Re$ and $n$ values. These contours provide essential information on the complicated pressure distribution within the three-dimensional lid-driven cavity flow, as impacted by different $Re$ and $n$. At $Re = 1$, the flow is primarily laminar, with pressure contours that follow a symmetrical distribution for all $n$ values. At $Re = 50$, the contours display asymmetry, unlike the $Re = 1$ case. The maximum pressure is seen around the top right and bottom corners. As the $Re$ increases to $Re = 100$ and $200$, the flow transitions to a turbulent phase, with distinct pressure contours depending on the power-law index $(n)$. 
At $Re = 100$, the increased pressure is concentrated on the right wall, but at $Re = 200$, it becomes more localized in the top and bottom corners next to the right wall $(x = 1)$ for all values of $n$. This phenomenon occurs because raising the $Re$ to 200 improves mixing and turbulence effects. We've noticed that raising the power-law index $(n)$ at a given Reynolds number $(Re)$ results in a higher value of maximum pressure. This phenomenon can be attributed to the rheological behavior of non-Newtonian fluids. Specifically, shear-thickening behaviour $(n = 1.5)$ causes increased viscosity and resistance to flow, resulting in increasing pressure accumulation. In contrast, shear-thinning behavior $(n = 0.5)$ leads to decreased viscosity and less resistance to flow, resulting in a smaller pressure buildup compared to Newtonian fluids. 
Hence, the pressure contours highlight the complex interaction between the power-law index $(n)$, $(Re)$, and flow behavior in a lid-driven cavity. Understanding these pressure patterns is critical for optimizing operations using non-Newtonian fluids and building systems that accommodate their rheological features.

\newpage
\section{Conclusion}
This paper introduces a novel higher-order super compact (HOSC) finite difference technique designed specifically for the non-Newtonian power-law model. Using this newly developed technique, we addressed the benchmark problem of 3D lid-driven cavity flow by carefully studying the varied rheological behaviors of shear-thinning ($n=0.5$), shear-thickening $(n=1.5)$, and Newtonian fluids across different Reynolds numbers $(Re= 1, 50, 100, 200)$. Prior to delving into the results and discussion, a thorough validation process was undertaken, comparing our findings with existing benchmark results for the power-law fluid model within a closed cavity. The outcome revealed an outstanding alignment with previously published results, affirming the exceptional consistency, accuracy, and reliability of our approach in capturing the intricate 3D phenomena inherent in the non-Newtonian power-law model. The findings of numerical computation lead to the following conclusions:
\begin{enumerate}
    \item The center of the primary vortex tends to shift towards the cavity's center with an increase of $n$ values at a given $Re$. Also, as $Re$ increases at a specific $n$, the center of the primary vortex tilts towards the right top corner of the cavity. However, this displacement is more pronounced for shear-thinning fluids ($n=0.5$).
    
    \item For shear-thinning fluid $(n=0.5)$, $u$ velocity is higher near the top moving wall than the case of Newtonian $(n=1.0)$ and shear-thickening fluid $(n=1.5)$ for all $Re$.
    
    \item As $Re$ increases for each specific $n$ value, higher gradients and increasing peak values for both $u$ and $w$ are observed. 

    \item Viscosity is low near the top moving wall for shear thinning fluid and viscosity is high near the moving lid for shear thickening fluid.
    \item High pressure is concentrated near the right wall for $Re = 1, 50$ and $100$, whereas for $Re = 200$, it becomes more localized near the top and bottom corners adjacent to the right wall $(x = 1)$ for all values of $n$.
\end{enumerate}
\vspace{11pt}
{\textbf{Author Declaration}}\\
The authors have no conflicts to disclose.
\vspace{11pt}\\
{\textbf{Data Availability}}\\
The data that support the findings of this study are available from the corresponding author upon reasonable request.

\end{document}